\documentclass[reprint, amsmath,amssymb, aps]{revtex4-2}

\usepackage[cp866]{inputenc}
\usepackage[english]{babel}
\usepackage{amsmath, amssymb}
\usepackage{graphicx}
\usepackage{amsthm}
\usepackage{epsfig}
\usepackage{bm}
\usepackage{indentfirst}

\newcommand{\be}{\begin{equation}}
 \newcommand{\ee}{\end{equation}}
 \newcommand{\bea}{\begin{eqnarray}}
 \newcommand{\eea}{\end{eqnarray}}

  \newcommand{\dst}{\displaystyle}
 \newcommand{\fr}[2]{\frac{{\dst #1}}{{\dst #2}}}

\usepackage[dvipsnames]{xcolor}

\def\ketm#1{  \left\vert  #1   \right\rangle   }

\def\sprm#1#2{  \left\langle #1 \left\vert \right. #2 \right\rangle   }

\def\mem#1#2#3{  \left\langle #1 \left\vert  #2 \right\vert #3 \right\rangle   }

\def\redmem#1#2#3{  \left\langle #1 \left\Vert
                  #2 \right\Vert #3 \right\rangle   }

\def\sixjm#1#2#3#4#5#6{  \left\{ \begin{array}{ccc}
                                               #1 & #2 & #3  \\
                                               #4 & #5 & #6
                     \end{array} \right\}   }
\begin{document}

%
% --------------------------- Title ----------------------------------------
%
\title{Resonant scattering of plane-wave and twisted photons at the Gamma Factory}

%
% ---------------------------- Authors -------------------------------------
%
\author{Valeriy~G.~Serbo}
\affiliation{Novosibirsk State University, RUS--630090, Novosibirsk, Russia}
\affiliation{Sobolev Institute of Mathematics, RUS--630090, Novosibirsk, Russia}

\author{Andrey Surzhykov}
\affiliation{Physikalisch--Technische Bundesanstalt, D--38116 Braunschweig, Germany}
\affiliation{Institut f\"ur Mathematische Physik, Technische Universit\"at Braunschweig, D--38106 Braunschweig, Germany}
\affiliation{Laboratory for Emerging Nanometrology Braunschweig, D-38106 Braunschweig, Germany}

\author{Andrey Volotka}
\affiliation{School of Physics and Engineering, ITMO University, RUS--199034, Saint-Petersburg, Russia}

\date{\today \\[0.3cm]}

%%%%%%%%%%%%%%%%%%%%%%%%%%%%%%%%%%%%%%%%%%%%%%%%%%%%%%%%%%%%%%%%%%%%%%%%
%
%======================       ABSTRACT         =========================
\begin{abstract}
We present a theoretical investigation of the resonant elastic scattering of laser photons by ultra--relativistic partially stripped ions, that is the core process of the Gamma Factory project. Special emphasis in our study is placed on the angular distribution and polarization of scattered photons as observed in the collider and ion--rest reference frames. In order to describe these (angular and polarization) properties for arbitrary relativistic many--electron ion, the general approach, based on the application of irreducible polarization tensors, is laid down. By making use of the polarization tensors we explore in detail the scattering of both, conventional plane--wave-- and twisted (or vortex) photons. For the former case we show how the propagation directions and polarization states of incident and outgoing photons are related to each other for the $n S_{0} \to n' P_{1} \to n S_{0}$, $n S_{1/2} \to n' P_{1/2} \to n S_{1/2}$ and $n S_{1/2} \to n' P_{3/2} \to n S_{1/2}$ resonant transitions. For the scattering of initially twisted light, that carries non--zero orbital angular momentum, we explore the angular distribution of secondary photons and discuss the conditions under which they are also twisted.
\end{abstract}
%======================    END ABSTRACT       ==========================
%
%\pacs{03.65.Pm, 34.80.Lx}
%%%%%%%%%%%%%%%%%%%%%%%%%%%%%%%%%%%%%%%%%%%%%%%%%%%%%%%%%%%%%%%%%%%%%%%%
\maketitle
%%%%%%%%%%%%%%%%%%%%%%%%%%%%%%%%%%%%%%%%%%%%%%%%%%%%%%%%%%%%%%%%%%%%%%%%

%
%
%
\section{Introduction}

A large number of studies in various branches of physics have been performed during the last years within the framework of the Physics Beyond Colliders initiative, whose goal is to further exploit the unique potential of the CERN facility \cite{JaL18}. One of the very promising projects, developed currently as a part of the initiative, is the Gamma Factory \cite{Kra15}. This project is located at the borderline between accelerator, atomic, nuclear and laser physics, and is focused on the production, storage and operation of partially stripped heavy ions in the CERN accelerator complex. In the heart of the Gamma Factory proposal is the fundamental process of the \textit{resonant photon scattering} in which a fast moving ion (i) is excited by a head--on incident photon and (ii) later decays with the emission of a secondary photon. Thanks to to the Lorentz transformation between the collider (laboratory) and ion--rest reference frames, this scattering process will lead to the production of high--energy and well--collimated secondary photon beams that can be used in a variety of experiments \cite{BuC20}.   

For the successful realization of the Gamma Factory project it is important to tune and control not only the energy (and intensity) of produced secondary photons but also their polarization and emission pattern. These---angular and polarization---properties are sensitive both to the shell structure of a ``target'' ion and to the propagation direction and polarization of incident light. Moreover, the application of twisted photon beams, which carry well--defined values of the orbital angular momentum (OAM) and has a helical phase front, can provide yet another opportunity for the ``fine--tuning'' of secondary radiation \cite{BuC20}. During the recent years, the twisted light has attracted a particular attention as a valuable tool in atomic and molecular physics \cite{BaB02,QuS17,FoA18,AfC18,AfC18b,BaA18}. Of special interest here is the \textit{production} of high--energy OAM photons, which can be achieved by using undulators and planar wigglers \cite{SaS08,BaH13,BoK20} as well as by Thomson and Compton scattering processes \cite{JeS11,JeS11b,TaK18,TaM18}. The resonant photon scattering, as will be realized at the Gamma Factory, may open up another promising route for the generation of twisted radiation in x-- or even gamma--ray domains.

One can conclude from the discussion above, that the guidance and analysis of the future studies at the Gamma Factory requires detailed knowledge about the \textit{interplay} of the atomic structure, geometry, polarization and even OAM--induced effects on the resonant scattering process. In this contribution, therefore, we lay down a general formalism for the unified description of these effects. We will focus, in particular, on the question of how the properties of incident primary light and of particular ionic transitions may affect the polarization and angular distribution of scattered photons if observed both in the collider and ion--rest frames. 

In order to present the theory of the resonant photon scattering we need to agree first about the geometry and kinematics of the process. They are briefly discussed in Sec.~\ref{sec:kinematic} for both (collider and ion--rest) reference frames and for the parameters of planned Gamma Factory experiments \cite{BuC20}. Later, in Sec.~\ref{subsec:relativistic_matrix_element} we apply the second--order perturbation theory for the electron--ion coupling in order to derive the scattering matrix element. This matrix element can be used to describe ``excitation--and--decay'' transition in an arbitrary relativistic many--electron ion and presents the \textit{building block} for evaluation of all the properties of scattering process. We demonstrate that the scattering matrix element can be written as a product of the (second--order) reduced amplitude, that reflects the electronic structure of an ion, and the so--called polarization tensor, that contains complete information about the geometry of the process and the polarization of photons. With help of the polarization--tensor approach we derive then in Sec.~\ref{subsec:amplitude_E1_} the matrix scattering element for the leading electric dipole (E1) scattering. In Sec.~\ref{subsec_relativistic_particular cases} this---dipole---matrix element is employed to investigate angular distribution and polarization of the photons, emitted in  $n S_{0} \to n' P_{1} \to n S_{0}$ and $n S_{1/2} \to n' P_{1/2, 3/2} \to n S_{1/2}$ transitions, which are of particular interest for the Gamma Factory research program. The non--relativistic limit of obtained results and their relation to the well--known Compton scattering data are discussed in Sec.~\ref{sec:non-relativistic_theory}. In this section, moreover, we present estimates for the \textit{total} cross sections of the resonant photon scattering. We argue, that for the representative parameters of the Gamma Factory these cross sections can reach peak values of approximately $10^{-13}$~cm$^2$, which is orders of magnitude larger than those of the backward Compton scattering.

While the first part of our paper is focused mainly on the analysis of the resonant scattering of conventional plane--wave photons, in Sec.~\ref{sec:scattering_twisted} we studied the head--on collisions between ion and twisted photon beams. For this case we again apply the polarization--tensor approach and investigate how the emission pattern of scattered photons can be affected by the kinematic parameters and angular momentum projection of twisted light. Moreover, in Sec.~\ref{subsec:TAM_transfer} we show that the ``twistedness'' can be even transferred---under particular conditions---from incoming (primary) to outgoing (secondary) photons. The summary of these important results and of the predictions for the incident plane--waves are given finally in Sec.~\ref{sec:summary}.

The present work is devoted to the development of a \textit{basis} for the description of the resonant photons scattering by partially stripped ions. It provides a set of theoretical tools that can be easily employed to calculate the polarization and angular properties of scattered photons for any setup of future Gamma Factory experiments. The analysis of such experiments will generally require also the use of the density matrix approach to account for a partial decoherence in the system ``ions + light''. This density matrix analysis, based on the present formulas, is carried out in the follow--up paper \cite{VoS21} to be published in this special issue.       

Relativistic units ( $\hbar = c = m_e = 1$) are used throughout the paper unless stated otherwise.

%
% ------------------------------------ Figure 1 -------------------------------------------
%
\begin{figure}[t]
	\includegraphics[width=0.95\linewidth]{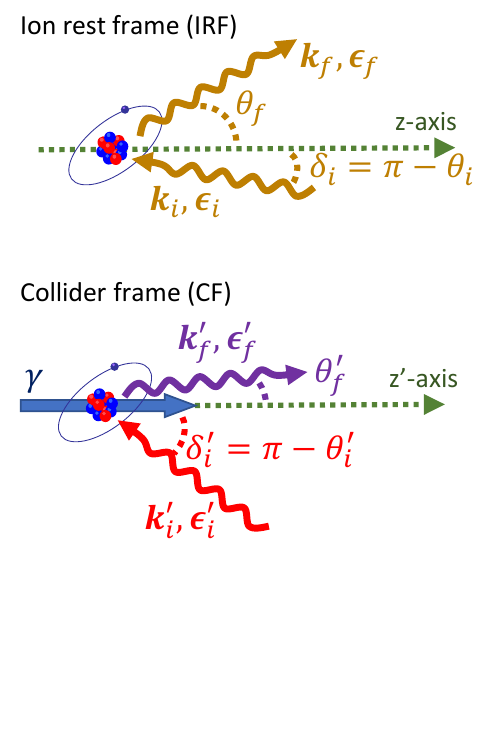}
	\vspace*{-3.0cm}
	\caption{The geometry of the photon scattering process in the ion rest frame (IRF) and collider frame (CF). In both frames the quantization ($z$--) axis is chosen along the ion's direction of propagation in the CF. Together with the $x$--axis (not displayed here), it defines the ($xz$--) plane of the collider ring. The propagation directions of incident and outgoing photons are defined by their polar, $\theta_{i,f}$, and azimuthal angles $\varphi_{i,f}$. While the latter are equal in the IRF and CF, $\varphi_{i} = \varphi'_{i}$ and $\varphi_{f} = \varphi'_{f}$, the polar angles are related to each other by Eq.~(\ref{eq:angle_transformation_exact}). Finally, since (almost) head--on laser--ion collisions are planned at the Gamma Factory, where $\theta_{i} \approx \pi$ and $\theta'_{i} \approx \pi$, the so--called crossing angles $\delta_i = \pi - \theta_i$ and $\delta'_i = \pi - \theta'_i$ will be also used to characterize incident photons.}
    \label{Fig1}
\end{figure}
%
% ------------------------------------------------------------------------------------------
%

%
% ------------------------------- Kinematic parameters --------------------------------------------
%
\section{Kinematic parameters and geometry of the scattering}
\label{sec:kinematic}

In order to investigate the resonant scattering of photons by (moving) partially stripped ions, we have to discuss first the kinematic parameters and the geometry of the process. This discussion depends critically on the choice of a reference \textit{frame}, in which the properties of incoming and outgoing photons are defined. Two frames are used in the present study: (i) while the theoretical analysis is performed most conveniently in the ion rest frame (IRF), $x y z$, (ii) the obtained predictions have to be transformed to the collider frame (CF), $x' y' z'$, to be compared with experimental data. In both frames the quantization axes $z$ and $z'$ are coincide and are chosen along the momentum of an ion in the CF, see Fig.~\ref{Fig1}. Moreover, the axes $x$ and $x'$ are chosen in a such way that the $xz$ and $x'z'$ planes coincide with the collider plane. The energy and kinematic characteristics of the scattering process in IRF and CF are briefly reviewed below.   

\subsection{Ion rest frame (IRF)}

Within the ion rest frame, the initial (incident) and final (scattered) photons are characterized by their linear momenta ${\bm k}_i$ and ${\bm k}_f$ as well as by the polarization vectors ${\bm \epsilon}_i$ and ${\bm \epsilon}_f$, respectively. The photon propagation directions are defined most naturally by the unit vectors ${\hat {\bm k}}_i = {\bm k}_i/k_i = \left(\sin\theta_i \cos\varphi_i, \sin\theta_i \sin\varphi_i, \cos\theta_i \right)$ and ${\hat {\bm k}}_f = {\bm k}_f/k_f = \left(\sin\theta_f \cos\varphi_f, \sin\theta_f \sin\varphi_f, \cos\theta_f \right)$, where $\theta_{i,f}$ and $\varphi_{i,f}$ are the polar and azimuthal angles in the IRF. Moreover, since in the Gamma Factory setup one investigates \textit{near} head--on ion--laser collisions, for which $\theta_i \approx \pi$, it is practical to describe the incoming photon by the so--called crossing angle $\delta_i = \pi - \theta_i$, c.~f.~Fig.~\ref{Fig1}. As we will see it later, this angle is usually very small in the IRF for the typical Gamma Factory energies.   

Beside the direction of propagation, the momenta ${\bm k}_i$ and ${\bm k}_f$ define also the energies of the photons, $\omega_i = k_i = \left|{\bm k}_i\right|$ and $\omega_f = k_f = \left|{\bm k}_f\right|$. Within the IRF and for the resonant elastic scattering, these energies are equal and, moreover, close to the transition energy between particular ionic levels with energies $E_i$ and $E_\nu$:
\begin{equation}
    \label{eq:energies_IRF}
    \omega_i = \omega_f \approx E_{\nu i} \equiv E_{\nu} - E_{i} \, .
\end{equation}
In Table~\ref{tab1} we listed several transitions in partially stripped ions, that attract interest in the Gamma Factory project. As seen from the table, these are x--ray transitions which proceed, moreover, between fine--structure states. It clearly justifies the use of the \textit{relativistic} theory for the description of the resonant photon scattering.

%
% ------------------------------ Table 1 --------------------------------------------------
%
\begin{table*}[t]
\begin{tabular}{llllll}
   \hline\\[-0.3cm]
    Ion \hspace*{0.8cm} & Transition \hspace*{2.7cm} & $E_{\nu i}$ [eV] \hspace*{0.7cm} & $\gamma$  \hspace*{1.0cm} & $\omega'_i$ [eV] \hspace*{0.3cm} & $(\omega_f')_{\max}$ [eV] \\[0.2cm]
   \hline
   Ar$^{16+}$ & 1s$^2$$\; {}^1S_0$ -- 1s 2p$\; {}^1P_1$       & 3139.6  & 96.3   & 16.3   & 6.0 $\times$ 10$^{5}$ \\
              &                                     &         & 2800   & 0.6  & 1.8 $\times$ 10$^{7}$ \\[0.2cm]    
   Xe$^{52+}$ & 1s$^2$$\; {}^1S_0$ -- 1s 2p$\; {}^1P_1$       & 30629.7 & 96.3   & 159.0   & 5.9 $\times$ 10$^{6}$ \\
              &                                     &         & 2800   & 5.5  & 1.7 $\times$ 10$^{8}$ \\[0.2cm]    
   Pb$^{79+}$ & 1s$^2$ 2s$\; {}^2S_{1/2}$ -- 1s$^2$ 2p$\; {}^2P_{1/2}$          & 230.8   & 96.3   & 1.2   & 4.4 $\times$ 10$^{4}$ \\
              &                                     &         & 2800   & 0.04  & 1.3 $\times$ 10$^{6}$ \\[0.2cm]    
   Pb$^{79+}$ & 1s$^2$ 2s$\; {}^2S_{1/2}$ -- 1s$^2$ 2p$\; {}^2P_{3/2}$      & 2642.2  & 96.3  & 13.7  & 5.1 $\times$ 10$^{5}$ \\
              &                                     &         & 2800  & 0.5   & 1.5 $\times$ 10$^{7}$ \\[0.2cm]        
   U$^{90+}$  & 1s$^2$$\; {}^1S_0$ -- 1s 2p$\; {}^1P_1$       & 100611  & 96.3  & 522.4 & 1.9 $\times$ 10$^{7}$ \\
              &                                     &         & 2800  & 18.0  & 5.6 $\times$ 10$^{8}$ \\      
   \hline
\end{tabular}
\caption{Bound--state transitions in partially stripped ions that are discussed in light of the planned experiments at the Gamma Factory \cite{BuC20}. Apart of the transition energy $E_{\nu i}$ as measured in the ion rest frame \cite{NIST_ASD, BuC20}, the energies $\omega'_i$ and $(\omega_f')_{\max}$ of the incident and elastically scattered photons in the collider (or laboratory) frame are presented. The latter two energies are obtained from Eqs.~(\ref{eq:energy_initial_transformation}) and (\ref{eq:energy_final}) for two relativistic Lorentz factors $\gamma$~=~96.3 and 2800, that correspond to the energies of ion beam in the proof--of--principle SPS and future LHC experiments, respectively.}
\label{tab1}
\end{table*}
\subsection{Collider frame (CF)}

As mentioned already above, the guidance and analysis of future experiments at the Gamma Factory will require the Lorentz transformation of the theoretical predictions, obtained in the IRF, into the collider (or laboratory) frame. In this frame, the momenta and polarization vectors of incident and scattered photons are denoted as $\left({\bm k}'_i,  {\bm \epsilon}'_i \right)$ and  $({\bm k}'_f,  {\bm \epsilon}'_f )$, respectively. Moreover, the polar $\theta'_{i,f}$ and azimutal $\varphi'_{i,f}$ are again employed to describe the propagation directions of the photons. While the azimutal angles do not change under the transformation between the reference frames, $\varphi'_{i,f} = \varphi_{i,f}$, the polar ones are connected by the standard relation:
\begin{equation}
    \cos\theta_{i,f} = \frac{\cos\theta'_{i,f} - v}{1 - v \cos\theta'_{i,f}} \, ,
    \label{eq:angle_transformation_exact}
\end{equation}
where $v$ is the velocity of an ion, which is directly related to the relativistic Lorentz factor $\gamma = 1 / \sqrt{1 - v^2}$. 

With the help of the general formula (\ref{eq:angle_transformation_exact}) one can further investigate the Lorentz transformation of the photon angles for the ultra--relativistic regime $\gamma \gg 1$, typical for the Gamma Factory facility. For example, most of the scattered (final--state) photons will be emitted into a very narrow forward angular cone:
\begin{equation}
    \cos\theta_f \approx \frac{1- (\gamma\theta'_f)^2}{1+(\gamma\theta'_f)^2} \, ,
    \label{eq:angle_transformation_final}
\end{equation}
which immediately implies the transformation relation
\begin{equation}
    \frac{d\Omega_f}{d\Omega'_f} \approx \frac{4\gamma^2}{[1+(\gamma \theta'_f)^2]^2} \, ,
    \label{eq:solid_angle_transformation}
\end{equation}
for the elements of the solid angle. Below these formulas will be used to relate the angle--differential scattering cross sections in ion--rest and collider frames. 

We can apply Eq.~(\ref{eq:angle_transformation_exact}) also to better understand the properties of the incident photons. As we mentioned already above, it is convenient to describe their propagation direction with respect to the ion beam axis in terms of the crossing angle $\delta'_i = \pi - \theta'_i$, with $0 \leq \delta'_i < \pi/2$ in the CF, see Fig.~\ref{Fig1}. The Lorentz transformation of this angle reads as:
\begin{equation}
    \cos\delta_i = \frac{\cos\delta'_i+v}{1+v \cos\delta'_i} \approx 1-\frac{\tan^2(\delta'_i/2)}{2\gamma^2} \, ,
    \label{eq:alpha_transformation_1}
\end{equation}
where the right--hand part is obtained again under assumption $\gamma \gg 1$. In this ultra--relativistic regime one can further evaluate Eq.~(\ref{eq:alpha_transformation_1}) and to find:
\begin{equation}
    \delta_i \approx \frac{\tan(\delta'_i/2)}{\gamma}< \fr{1}{\gamma} \, , 
    \label{eq:alpha_transformation_2}
\end{equation}
which implies that---owing to the Lorentz transformation---the crossing angle $\delta_i$ in the IRF does not exceed $0.6$~deg for the typical SPS and LHC energies. Moreover, for $\delta'_i \sim  $~1--2~deg in the collider frame, as planned for the Gamma Factory experiments with conventional plane--wave radiation, we obtain:
\begin{equation}
    \delta_i \approx \frac{\delta'_i}{2\gamma} \, ,
    \label{eq:alpha_transformation_3}
\end{equation}
i.e. even stronger restriction on the IRF angle $\delta_i$.

The Lorentz transformation between the reference frames strongly influences not only the propagation directions but also the energies of the incident and outgoing photons. Indeed, in contrast to the simple relation (\ref{eq:energies_IRF}), valid for the IRF, the energies $\omega'_i$ and $\omega'_f$ \textit{substantially} differ from each other in the CF. For example, a laser with the energy:
\begin{equation}
    \omega'_i \approx \frac{E_{\nu i}}{\gamma (1+\cos\delta'_i)} \approx \frac{E_{\nu i}}{2\gamma} \, , 
    \label{eq:energy_initial_transformation}
\end{equation}
has to be installed in the collider frame in order to (nearly) match the energy $E_{\nu i} = E_{\nu} - E_i$ of a bound--state transition in a ion, moving with the Lorentz factor $\gamma$. Much higher energy,
\begin{equation}
    \label{eq:energy_final}
    \left(\omega'_f\right)_{\rm max} = 2\gamma E_{\nu i} \approx (2\gamma)^2 \omega'_i \, ,
\end{equation}
will be brought away by the back--scattered photon that propagates along the ion beam axis, $\theta'_f = 0$. The values of $\omega'_i$ and $(\omega'_f)_{\rm max}$ for the typical Gamma Factory parameters are presented in Table~\ref{tab1}.

%
% ---------------------------- Section: Relativistic theory --------------------------------------
%
\section{Plane--wave scattering: Relativistic theory}
\label{sec:relativistic_theory}

\subsection{Scattering matrix element}
\label{subsec:relativistic_matrix_element}

Not much has to be said about the general approach to describe scattering of photons by many--electron ions. Within this approach, that has been intensively discussed in the literature \cite{KiP85,KaK86,RoK99}, the interaction between electromagnetic radiation and atomic electrons is treated by means of the second--order perturbation theory. Hence, the scattering cross section as well as all the properties of the outgoing  photons can be expressed in terms of the (second--order) matrix element. For the highly--energetic photons and heavy ions, the \textit{relativistic} form of this matrix element has to be employed:
\begin{widetext}
\begin{eqnarray}
   \label{eq:amplitude_general}
   {\cal M}^{\rm (pl)}_{M_f, M_i} &=& \alpha \,
   \sum\limits_{\alpha_\nu J_\nu M_\nu}
  \left[ \frac{\mem{f}{\hat{\mathcal{R}}^\dag({\bm k}_f, {\bm \epsilon}_f)
                              }{\nu} \:
         \mem{\nu}{\hat{\mathcal{R}}({\bm k}_i, {\bm \epsilon}_i)
                              }{i}}{E_i - E_\nu + \omega_i}+
   \frac{\mem{f}{\hat{\mathcal{R}}({\bm k}_i, {\bm \epsilon}_i)
                              }{\nu} \:
         \mem{\nu}{\hat{\mathcal{R}}^\dag({\bm k}_f, {\bm \epsilon}_f)
                              }{i}}{E_i - E_\nu - \omega_i}
                              \right]\, ,
\end{eqnarray}
\end{widetext}
where $\ketm{i} \equiv \ketm{\alpha_i J_i M_i}$ and $\ketm{f} \equiv \ketm{\alpha_f J_f M_f}$ denote (many-electron) states of an ion just before and after the scattering has occurred, while $\ketm{\nu} \equiv \ketm{\alpha_\nu J_\nu M_\nu}$ is the intermediate state. Apart from the total angular momenta $J_{i,f}$ and their projections $M_{i,f}$, the $\alpha_{i,f}$ refers to additional quantum numbers that are needed for a unique specification of these states. Below we will apply Eq.~(\ref{eq:amplitude_general}) to the particular case of the \textit{elastic} scattering, for which $E_i = E_f$, $\alpha_i = \alpha_f$ and $J_i = J_f$. 

To further evaluate the matrix element (\ref{eq:amplitude_general}) we have to discuss the operators $\hat{\mathcal{R}}({\bm k}_i, {\bm \epsilon}_i)$ and $\hat{\mathcal{R}}^{\dag}({\bm k}_f, {\bm \epsilon}_f)$ that describe absorption and emission of photons with momentum and polarization vectors ${\bm k}_i, {\bm \epsilon}_i$ and ${\bm k}_f, {\bm \epsilon}_f$, respectively. As usual in atomic calculations, these many--particle operators can be written as a sum of their one--electron counterpartners. For example, the photon--absorption operator reads in the Coulomb gauge as:
\begin{eqnarray}
   \label{eq:R_operator_general}
   \hat{\mathcal{R}}(\bm{k}_i, \bm{\epsilon}_i) &=& 
   \sum_{m} {\bm \alpha}_m \cdot {\bm \epsilon}_i \, {\rm e}^{i {\bm k}_i {\bm r}_m} \, ,
\end{eqnarray}
where ${\bm r}_m$ is the coordinate of the $m$--th electrons, and ${\bm \alpha}_m = \left( \alpha_{x,m}, \alpha_{y,m}, \alpha_{z,m} \right)$ is the vector of the Dirac matrices. The evaluation of the matrix element (\ref{eq:amplitude_general}) can be significantly simplified if the one--particle operator ${\bm \alpha}_m \cdot {\bm \epsilon}_i \, {\rm e}^{i {\bm k}_i {\bm r}_m}$ from Eq.~(\ref{eq:R_operator_general}) is further expanded into partial--wave components which posses well--defined symmetry properties. For the propagation of a photon in the direction $\hat{{\bm k}}_i = {\bm k}_i/k_i$, characterized by polar and azimuthal angles $\theta_i$ and $\varphi_i$, this expansion reads as:
\begin{eqnarray}
   \label{eq:A_decomposition}
   {\bm \alpha}_m \cdot {\bm \epsilon}_i \, {\rm e}^{i {\bm k}_i {\bm r}_m} & = & 4 \pi \sum\limits_{p L M} i^{L-|p|} \,
   \left( \bm{\epsilon}_i \cdot \bm{Y}^{(p) *}_{L M}(\hat{\bm{k}}_i) \right) \, \nonumber \\[0.2cm]
   &  & \qquad\times \;
   \left( {\bm \alpha}_{m} \cdot \bm{a}^{(p)}_{L M}(\omega_i ; \, {\bm r}_m) \right)\, ,
\end{eqnarray}
with $\bm{Y}^{(p)}_{L M}(\hat{\bm{k}})$ being the vector spherical harmonics, and $\bm{a}^{p}_{L M}(\omega_i ; \, {\bm r}_m)$ -- electric ($p$~=~1) and magnetic ($p$~=~0) multipole components of the electromagnetic field. These components, whose explicit form can be found in Ref.~\cite{MaM00,VaM88}, are constructed to represent irreducible tensors of rank $L$. The use of the irreducible tensors allows one to apply the Wigner--Eckart theorem and, after some algebra, re--write the scattering matrix element as:

\begin{eqnarray}
   \label{eq:matrix_evaluated_2}
   {\cal M}^{\rm (pl)}_{M_f, M_i}
   &=& \sum\limits_{k q} \sqrt{2k + 1} \sprm{k q \, J_f M_f}{J_i M_i} \nonumber \\
   && \qquad \times U_{k \, q}(\alpha_f J_f; \, \alpha_i J_i) \, ,
\end{eqnarray}
where $\left| J_i - J_f\right| \le k \le J_i + J_f$ and the function $U_{kq}$ is given by:
\begin{widetext}
\begin{eqnarray}
   \label{eq:U_definition}
   U_{k \, q}(\alpha_f J_f; \, \alpha_i J_i)  &=& \frac{(4 \pi)^2 \, \alpha}{\sqrt{2J_i + 1}} \:
   \sum\limits_{L_1 p_1} \: \sum\limits_{L_2 p_2} i^{L_1 + |p_1| - L_2  - |p_2|} \,
   (-1)^{J_f + J_i} \, T_{k \, q}^{L_1 p_1; L_2 p_2} \;
   (\hat{\bm{k}}_i, \bm{\epsilon}_i \,; \hat{\bm{k}}_f, \bm{\epsilon}_f) \nonumber \\[0.2cm]
   && \hspace*{-1cm} \times \sum\limits_{J_{\nu}} \Bigg[ \sixjm{L_2}{L_1}{k}{J_i}{J_f}{J_\nu}
   S^{J_\nu}_{L_2 p_2, L_1 p_1}(\omega_i)
    + (-1)^{L_1 + L_2 + k} \sixjm{L_2}{L_1}{k}{J_f}{J_i}{J_\nu}
   S^{J_\nu}_{L_1 p_1, L_2 p_2}(-\omega_i) \Bigg] \, .
\end{eqnarray}
As seen from this expression, the function $U_{kq}$ and, hence, the matrix element can be written as a product of their radial and angular parts. The radial part is represented by the \textit{reduced} matrix elements: 
\begin{eqnarray}
   \label{eq:S_function}
   S^{J_\nu}_{L_1 p_1, \, L_2 p_2}(\pm \, \omega_i)
   &=&
   \sum\limits_{\alpha_\nu} \frac{\redmem{\alpha_f J_f}{\sum\limits_{m} {\bm \alpha}_m \cdot
   \hat{\bm a}^{(p_1)}_{L_1}(\omega_i ; \, {\bm r}_m)}{\alpha_\nu J_\nu}
   \redmem{\alpha_\nu J_\nu}{\sum\limits_{m} {\bm \alpha}_m \cdot
   \hat{\bm a}^{(p_2)}_{L_2, m}(\omega_i ; \, {\bm r}_m)}{\alpha_i J_i}}{E_\nu - E_i \mp \omega_i} \, ,
\end{eqnarray}
that are independent on the polarization states and propagation directions of incident and outgoing photons as well as on the magnetic sublevel population of a target ion. In contrast, the angle and polarization dependence of the elastic scattering is defined by the irreducible tensor 
\begin{eqnarray}
   \label{eq:T_tensor}
   T_{k \, q}^{L_1 p_1; L_2 p_2}(\hat{\bm{k}}_i, \bm{\epsilon}_i ; \, \hat{\bm{k}}_f, \bm{\epsilon}_f)
   & = &
   \left\{ \left( \bm{\epsilon}_i \cdot \bm{Y}^{(p_1)}_{L_1}(\hat{\bm{k}}_i) \right)  \otimes
           \left( \bm{\epsilon}^{*}_f \cdot \bm{Y}^{(p_2)}_{L_2}(\hat{\bm{k}}_f) \right)
   \right\}_{k \, q} \nonumber \\[0.2cm]
   & = & \sum\limits_{M_1 M_2} \sprm{L_1 M_1 \, L_2 M_2}{k q}
   \left( \bm{\epsilon}_i \cdot \bm{Y}^{(p_1)}_{L_1 M_1}(\hat{\bm{k}}_i) \right) \:
   \left( \bm{\epsilon}^{*}_f \cdot \bm{Y}^{(p_2)}_{L_2 M_2}(\hat{\bm{k}}_f) \right) \, ,
\end{eqnarray}
which, in turn, is insensitive to the electronic structure of a particular ion, but reflects the geometry of the process. While the general properties of this so--called polarization tensor have been discussed in detail by Manakov and co-workers \cite{MaM00,MaM01}, below we will present the explicit form of $T_{k \, q}^{L_1 p_1; L_2 p_2}$ for the case of the leading electric dipole approximation for which $L_1 = L_2 = 1$ and $p_1 = p_2 = 1$.

\subsection{Resonant $E1$ approximation}
\label{subsec:amplitude_E1_}

Equations~(\ref{eq:matrix_evaluated_2})--(\ref{eq:T_tensor}) provide the most general form of the matrix element ${\cal M}^{\rm (pl)}_{M_f, M_i}$, that can be used to describe the elastic scattering of a photon of an arbitrary energy $\omega_i$ and via all allowed multipole channels $\left(L_1 p_1, \; L_2 p_2 \right)$. In the present work, however, we restrict ourselves to the case of the \textit{resonant} scattering:
\begin{equation}
    \label{eq:scheme_process}
    \ketm{\alpha_i J_i} +\gamma_i({\bm k}_i {\bm \epsilon}_i) \to \ketm{\alpha_\nu J_\nu} \to \ketm{\alpha_i J_i} + \gamma_f({\bm k}_f {\bm \epsilon}_f) \, ,
\end{equation}
which takes place when the photon energy is very close to the excitation energy of some ionic state, $\omega_i \approx E_{\nu} - E_i$. This particular state, therefore, gives the dominant contribution to the intermediate--state summation in the second--order amplitude, thus allowing us to simplify the reduced matrix element to:
\begin{eqnarray}
   \label{eq:S_function_E1}
   S^{J_\nu, {\rm res}}_{L_1 p_1, \, L_2 p_2}(\omega_i)
   &=&
   \frac{\redmem{\alpha_i J_i}{\sum\limits_{m} {\bm \alpha}_m \,
   \hat{\bm a}^{(p_1)}_{L_1, m}(k)}{\alpha_\nu J_\nu}
   \redmem{\alpha_\nu J_\nu}{\sum\limits_{m} {\bm \alpha}_m \,
   \hat{\bm a}^{(p_2)}_{L_2, m}(k)}{\alpha_i J_i}}{E_\nu - E_i - \omega_i - i\Gamma_{\nu}/2} \, ,
\end{eqnarray}
where the width $\Gamma_{\nu}$ of the excited state $\ketm{\alpha_\nu J_\nu}$ is introduced in the denominator. The validity of this so--called resonant approximation has been discussed in detail in Ref.~\cite{SaV20}. 

If the resonant scattering (\ref{eq:scheme_process}) proceeds, moreover, via absorption and emission of electric dipole ($E1$) photons, the further evaluation of the function $U_{kq}$ is also possible and leads to: 
\begin{eqnarray}
   \label{eq:U_definition_E1}
   U^{(E1, {\rm res})}_{k \, q}(\alpha_i J_i; \alpha_i J_i)  &=& \frac{(4 \pi)^2 \, \alpha}{\sqrt{2J_i + 1}} \:
   (-1)^{2 J_i} \, \sixjm{1}{1}{k}{J_i}{J_i}{J_\nu}  \, T_{k \, q}^{E1; E1}(\hat{\bm{k}}_i, \bm{\epsilon}_i \,; \hat{\bm{k}}_f, \bm{\epsilon}_f) \,
   S^{J_\nu, {\rm res}}_{E1, E1}(\omega_i) \, ,
\end{eqnarray}
where we used the short--hand notations $S^{J_\nu, {\rm res}}_{E1, E1}(\omega_i) \equiv S^{J_\nu, {\rm res}}_{1 1, \, 1 1}(\omega_i)$ and $T_{k \, q}^{E1; E1} \equiv T_{k \, q}^{1 1; \, 1 1}$ for the multipole pair $(L_1 = 1, \, p_1 = 1)$---$(L_2 = 1, \, p_2 = 1)$. By inserting the function (\ref{eq:U_definition_E1}) into Eq.~(\ref{eq:matrix_evaluated_2}) we finally obtain the scattering matrix element within the resonant dipole approximation:
\begin{eqnarray}
    \label{eq:matrix_evaluated_E1}
    {\cal M}^{({\rm pl}, E1)}_{M_f, M_i} &=& \frac{(4 \pi)^2 \, \alpha}{\sqrt{2J_i + 1}} \: (-1)^{2 J_i} \:
    S^{J_\nu, {\rm res}}_{E1, E1}(\omega_i) \nonumber \\
    &\times& \sum_{kq} (-1)^k \, \sqrt{2k + 1} \, \sprm{k q \, J_i M_f}{J_i M_i} \,  \sixjm{1}{1}{k}{J_i}{J_i}{J_\nu} \,  T_{k \, q}^{E1; E1}(\hat{\bm{k}}_i, \bm{\epsilon}_i \,; \hat{\bm{k}}_f, \bm{\epsilon}_f) \, .
\end{eqnarray}
\end{widetext}
As mentioned already above, all the information about the nuclear charge $Z$ and shell structure of a particular ion is contained exclusively in the reduced matrix element $S^{J_\nu, {\rm res}}_{E1, \,E1}(\omega_i)$, which is just a pre--factor in the first line of Eq.~(\ref{eq:matrix_evaluated_E1}). Therefore, the angular and polarization properties of the elastically scattered photons will depend only on symmetry of initial (final) and intermediate ionic states $\ketm{\alpha_i J_i}$ and $\ketm{\alpha_\nu J_\nu}$, as well as on the propagation direction $\hat{{\bm k}}_i$ and polarization ${\bm \epsilon}_i$ of incident light. Beside the trivial Wigner symbols, this dependence arises from the tensor 
\begin{eqnarray}
    \label{eq:T_tensor_E1}
    T_{k \, q}^{E1; E1}(\hat{\bm{k}}_i, \bm{\epsilon}_i \,; \hat{\bm{k}}_f, \bm{\epsilon}_f) 
    = \frac{3}{8\pi} \{ \bm{\epsilon}^*_f \otimes \bm{\epsilon}_i \}_{kq} \, ,
\end{eqnarray}
which can be written, within the $E1$ approximation, as the tensor product of polarization vectors of incident and outgoing photons. 

\subsubsection{Polarization tensor $T_{k \, q}^{E1; E1}$}
\label{subsubsec_tensor_T}

For the analysis of the angular and polarization properties of the scattered photons it might be useful to evaluate the polarization tensor (\ref{eq:T_tensor_E1}) even further. This evaluation would require the knowledge about the rank $k$ of the $T_{k \, q}^{E1; E1}$. As we will see below, the tensors of \textit{zero} and \textit{first} rank are needed to describe the resonant transitions that are likely to be studied in the first experiments at the Gamma Factory. These tensors can be expressed as scalar and vector product of polarization vectors as:
\begin{subequations}
\begin{equation}
\label{eq:T00_vectors}
    T_{0 \, 0}^{E1; E1}(\hat{\bm{k}}_i, \bm{\epsilon}_i \,; \hat{\bm{k}}_f, \bm{\epsilon}_f)  =  
    - \frac{\sqrt{3}}{8\pi} \, (\bm{\epsilon}^*_f \cdot \bm{\epsilon}_{i}) \, , 
\end{equation}
\begin{equation}
    T_{1 \, q}^{E1; E1}(\hat{\bm{k}}_i, \bm{\epsilon}_i \,; \hat{\bm{k}}_f, \bm{\epsilon}_f) =  \, 
     \frac{3i}{8\sqrt{2}\pi} \, \left[ \bm{\epsilon}^*_f \times \bm{\epsilon}_{i} \right]_q \, ,
     \label{eq:T10_vectors}
\end{equation}
\end{subequations}
where in the second expression $q = 0, \pm 1$ and $[...]_q$ is understood as the $q$--th vector component in the spherical coordinate system. 

Eqs.~(\ref{eq:T00_vectors})--(\ref{eq:T10_vectors}) allow one to calculate the tensors $T_{k \, q}^{E1; E1}$ for an \textit{arbitrary} polarization vectors of incident and outgoing photons. In the discussions below, however, we will often employ the \textit{circular} polarization basis ${\bm \epsilon}_{\lambda}$ with $\lambda = \pm 1$ being helicity, i.e. the spin projection of a photon on its propagation direction $\hat{{\bm k}}$. The vectors ${\bm \epsilon}_{\lambda}$ can be expanded in terms of spherical basis unit vectors ${\bm e}_{\sigma}$ as:
\begin{equation}
    \label{eq:polarization_vectore_helicity_basis}
    {\bm \epsilon}_{\lambda} = \sum_{\sigma=0,\,\pm 1} e^{-i \sigma \varphi} \, d^{\;1}_{\sigma \lambda}(\theta) \, {\bm e}_{\sigma} \, ,
\end{equation}
where $d^{\;1}_{\sigma \lambda}$ are elements of the small Wigner $d$--matrix and the angles $(\theta, \varphi)$ define the direction of the photon momentum ${\bm k}$. By using this expression one can easily evaluate both tensors (\ref{eq:T00_vectors})--(\ref{eq:T10_vectors}) for the \textit{circular} polarization vectors. For example, the zero--rank tensor reads as:
\begin{eqnarray}
   \label{eq:T00_helicity}
   T_{0 \, 0}^{E1; E1}(\hat{\bm{k}}_i, \bm{\epsilon}_{\lambda_i} \,; \hat{\bm{k}}_f, \bm{\epsilon}_{\lambda_f})  &=& - \frac{\sqrt{3}}{8\pi} \, (\bm{\epsilon}^*_{\lambda_f} \cdot \bm{\epsilon}_{\lambda_i}) \nonumber \\[0.2cm]
   && \hspace{-3.4cm} = - \frac{\sqrt{3}}{8\pi} \, \sum_{\sigma=0, \pm 1} \, {\rm e}^{i\sigma(\varphi_f-\varphi_i)} \,
   d^{\;1}_{\sigma\lambda_i}(\theta_i) \, d^{\;1}_{\sigma\lambda_f}(\theta_f) \, ,
\end{eqnarray}
and a bit more complicated formula can be obtained for the $T_{1 \, q}^{E1; E1}(\hat{\bm{k}}_i, \bm{\epsilon}_{\lambda_i} \,; \hat{\bm{k}}_f, \bm{\epsilon}_{\lambda_f})$.  

\subsection{Differential cross sections and polarization properties for particular cases}
\label{subsec_relativistic_particular cases}

With the help of the matrix element (\ref{eq:matrix_evaluated_E1}) one can calculate both the angular distribution and the polarization of the resonantly scattered photons. As mentioned already above, these angular and polarization properties depend on the choice of initial and intermediate states of a target ion. In what follows, therefore, we will consider particular examples of $n S_{0} \to n' P_{1} \to n S_{0}$ as well as $n S_{1/2} \to n' P_{1/2, 3/2} \to n S_{1/2}$ transitions which are of interest for the planned experiments at the Gamma Factory.

\subsubsection{$n S_{0} \to n' P_{1} \to n S_{0}$ scattering}

First, we will discuss the elastic scattering of photons by an ion (or atom) being in the $n S_{0}$ state for which $J_i = J_f = 0$ and $M_i = M_f = 0$. We assume, moreover, that the incident photon energy is chosen in such a way that the scattering proceeds via intermediate $n' P_{1}$ state.
In this case, summation over $k$ in Eq.~(\ref{eq:matrix_evaluated_E1}) is restricted to just a single term $k=0$ and the matrix element simplifies to 
\begin{eqnarray}
   \label{eq:matrix_evaluated_E1_0_to_0}
   {\cal M}^{({\rm pl}, E1)}(n S_{0} \to n' P_{1}) &\equiv& {\cal M}^{({\rm pl}, E1)}_{M_f = 0, M_i = 0}(n S_{0} \to n' P_{1})  \nonumber \\[0.2cm]
   && \hspace*{-1cm} = -2\pi \, \alpha \, S^{J_\nu = 1}_{E1, \,E1}(\omega_i) \, (\bm{\epsilon}^*_{f} \cdot \bm{\epsilon}_i) \, ,
\end{eqnarray}
where we employed Eq.~(\ref{eq:T00_vectors}) for the zero--rank polarization tensor. It worth stressing that the derived matrix element is exact for \textit{any} system that proceeds the resonant $n S_{0} \to n' P_{1} \to n S_{0}$ photon scattering. Moreover, by neglecting the spin effects and characterizing ionic states by their orbital angular momentum $L$ instead of the total one $J$, the same expression can be obtained for the non--relativistic $nS \to n'P \to nS$ transition. This non--relativistic limit of the theory will be discussed later in Sec.~\ref{subsec:non-rel_theory_resonant scattering} for particular case of hydrogen--like ions.

With the help of the matrix element (\ref{eq:matrix_evaluated_E1_0_to_0}) one can calculate both the angular distribution and polarization of the scattered photons. For instance, the angle--differential scattering cross section is obtained as:
\begin{eqnarray}
    \label{eq:diff_cross_section_0_to_0}
    \frac{{\rm d}\sigma^{\rm (pl)}}{{\rm d}\Omega_f} &=& \left| {\cal M}^{({\rm pl}, E1)}(n S_{0} \to n' P_{1}) \right|^2 \nonumber \\[0.2cm]
    &=& \left|R_{J_\nu = 1}(\omega_i)\right|^2 \, \left|(\bm \epsilon_f^* \cdot {\bm \epsilon}_i)\right|^2 \, ,
\end{eqnarray}
where the function $R_{J_\nu}(\omega_i)$ depends only on the energies of the incident photon $\omega_i$ and of the transition $E_{\nu i} = E_{\nu} - E_i$ as well as on the width of the excited state:
\begin{eqnarray}
   \label{eq:R_function_definition}
   R_{J_\nu}(\omega_i) &=& 2 \pi \alpha \, \frac{\left|\redmem{\alpha_\nu J_\nu}{\sum\limits_{m} {\bm \alpha}_m \,
   \hat{\bm a}^{(1)}_{1}(\omega_i; \, {\bm r}_m)}{\alpha_i J_i}\right|^2}{E_{\nu i} - \omega_i - i \Gamma_{\nu}/2}
   \nonumber \\[0.2cm]
   &=& \frac{2J_\nu + 1}{2 \omega_i} \, \frac{\Gamma_{\nu i}}/2{E_{\nu i} - \omega_i - i \Gamma_{\nu}/2} \, .
\end{eqnarray}
In the second line of this expression we used the standard relation between the (square of the) reduced matrix element and the partial width $\Gamma_{\nu i}$ for the radiative decay $\ketm{\nu} \to \ketm{i}$.  

The angle--differential cross section (\ref{eq:diff_cross_section_0_to_0}) is still sensitive to the polarization vectors of the incident and outgoing photons. However, it can be also used to investigate the scenario in which incident radiation is unpolarized and the scattered photons are detected by polarization--insensitive detectors. In order to derive the angle--differential cross section for this case, we first average over the incident--state and to sum over the final--state polarizations:
\begin{equation}
    \label{eq:polarization_summation}
    \frac{1}{2} \sum\limits_{\epsilon_i \epsilon_f} \left| \bm \epsilon_f^* \cdot {\bm \epsilon}_i \right|^2 = \frac{1}{2} \, \left(1 + \cos^2\theta_f \right) \, ,
\end{equation}
and, after inserting this expression into Eq.~(\ref{eq:diff_cross_section_0_to_0}), obtain:
\begin{equation}
    \label{eq:diff_cross_section_0_to_0_unpolarized}
    \frac{{\rm d}\sigma^{\rm (pl, unp)}}{{\rm d}\Omega_f} = \frac{1}{2}\left|R_{J_\nu = 1}(\omega_i)\right|^2 \, \left(1 + \cos^2\theta_f \right) \, .
\end{equation}
Here we assumed, moreover, the head--on collisions, for which $\theta_i = \pi$ and the polar angle $\theta_f$ describes the emission direction of the scattered photon with respect to the ion beam axis, chosen as the quantization axis. Upon the integration over this angle one can finally obtain the total cross section:
\begin{equation}
    \label{eq:total_cross_section_0_to_0_unpolarized}  
    \sigma^{\rm (pl, unp)} = \frac{8 \pi}{3} \, \left|R_{J_\nu = 1}(\omega_i)\right|^2 \, ,
\end{equation}
that depends both on the natural width $\Gamma_{\nu}$ and on the detuning $E_{\nu i} - \omega_i$, c.\,f.~Eq.~(\ref{eq:R_function_definition}). 

%
% ---------------------------------------------- Fig 2 ---------------------------------------------
%
\begin{figure}[t]
    \includegraphics[width=0.99\linewidth]{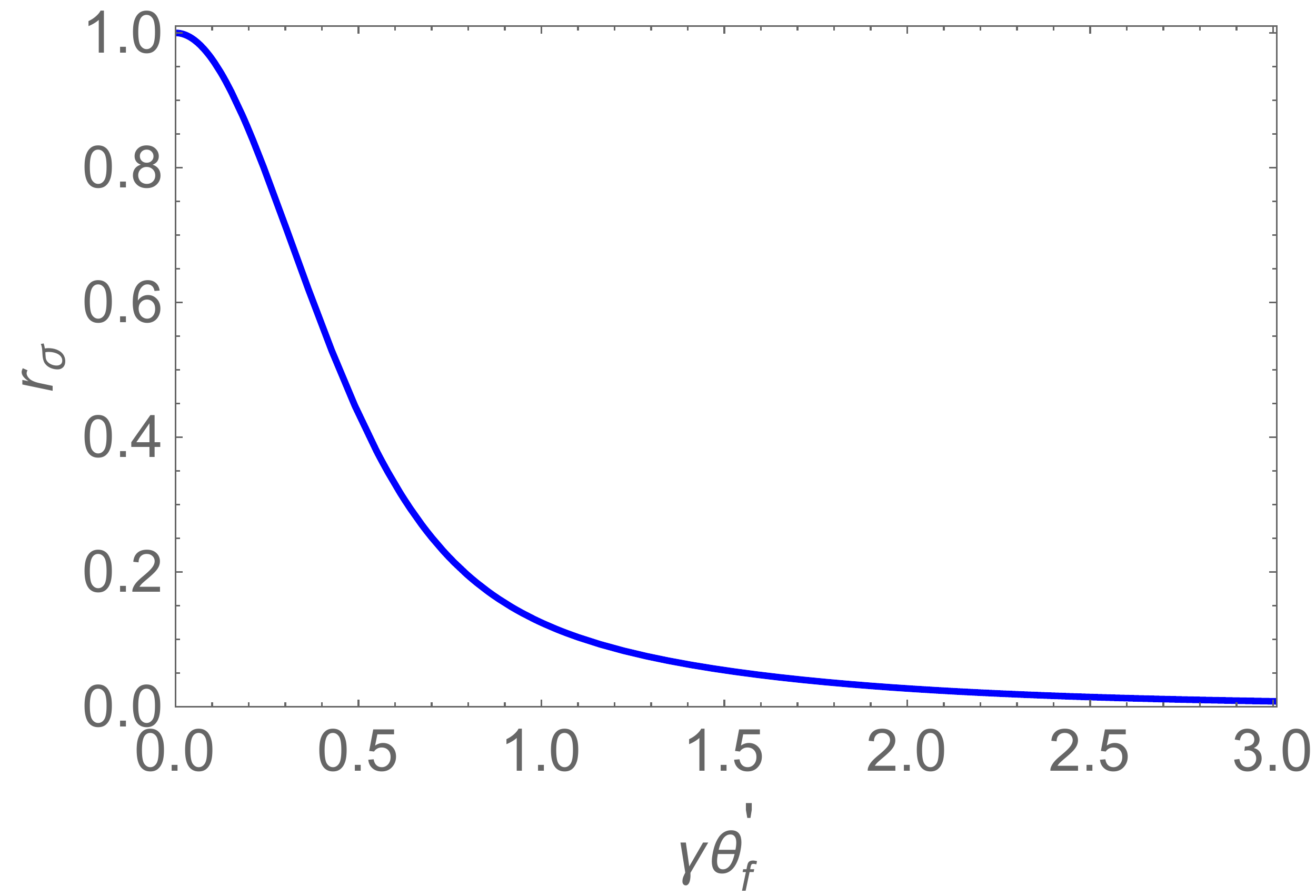}
	\caption{The angle--differential scattering cross section $r_{\sigma} = \fr{{\rm d}\sigma^{\rm (pl, unp)}(\theta_f')}{{\rm d}\Omega'_f}\Big/\fr{{\rm d}\sigma^{\rm (pl, unp)}(\theta_f'=0)}{{\rm d}\Omega'_f}$ normalized to the zero angle value. The cross section is calculated within the laboratory frame and for the $n S_{0} \to n' P_{1} \to n S_{0}$ resonant transition.}
    \label{Fig2}
\end{figure}
%
% --------------------------------------------------------------------------------------------------
%

Eq.~(\ref{eq:diff_cross_section_0_to_0_unpolarized}) describes the angle--differential scattering cross section in the rest frame of an ion. By employing basic theory from Sec.~\ref{sec:kinematic} one can easily convert ${\rm d}\sigma^{\rm (pl, unp)}/{\rm d}\Omega_f$ to the collider frame:
\begin{equation}
    \label{eq:diff_cross_section_0_to_0_unpolarized_lab_frame}
    \frac{{\rm d}\sigma^{\rm (pl, unp)}}{{\rm d}\Omega'_f} = 
    \left|R_{J_\nu = 1}(\omega_i)\right|^2 \, 
    4 \gamma^2\, \frac{1+(\gamma \theta'_f)^4}{\left[1+(\gamma \theta'_f)^2\right]^4} \, ,
\end{equation}
where the (CF) angle $\theta'_f$ is again defined with respect to the ion beam direction. Moreover, Eq.~(\ref{eq:diff_cross_section_0_to_0_unpolarized_lab_frame}) is obtained under the assumption that most of the scattered photons are emitted in a very narrow \textit{CF} angular range in the forward direction (relative to the ion beam) and, hence, $\theta'_f \lesssim 1/\gamma << 1$. This assumption is well supported by Fig.~\ref{Fig2} where we display the normalized cross section $R={\rm d}\sigma^{\rm (pl, unp)}(\theta'_f)/{\rm d}\sigma^{\rm (pl, unp)}(0)$ as the function of the parameter $\gamma \theta'_f$.

%
% ---------------------------------------------- Fig 3 ---------------------------------------------
%
\begin{figure}[t]
    \includegraphics[width=0.99\linewidth]{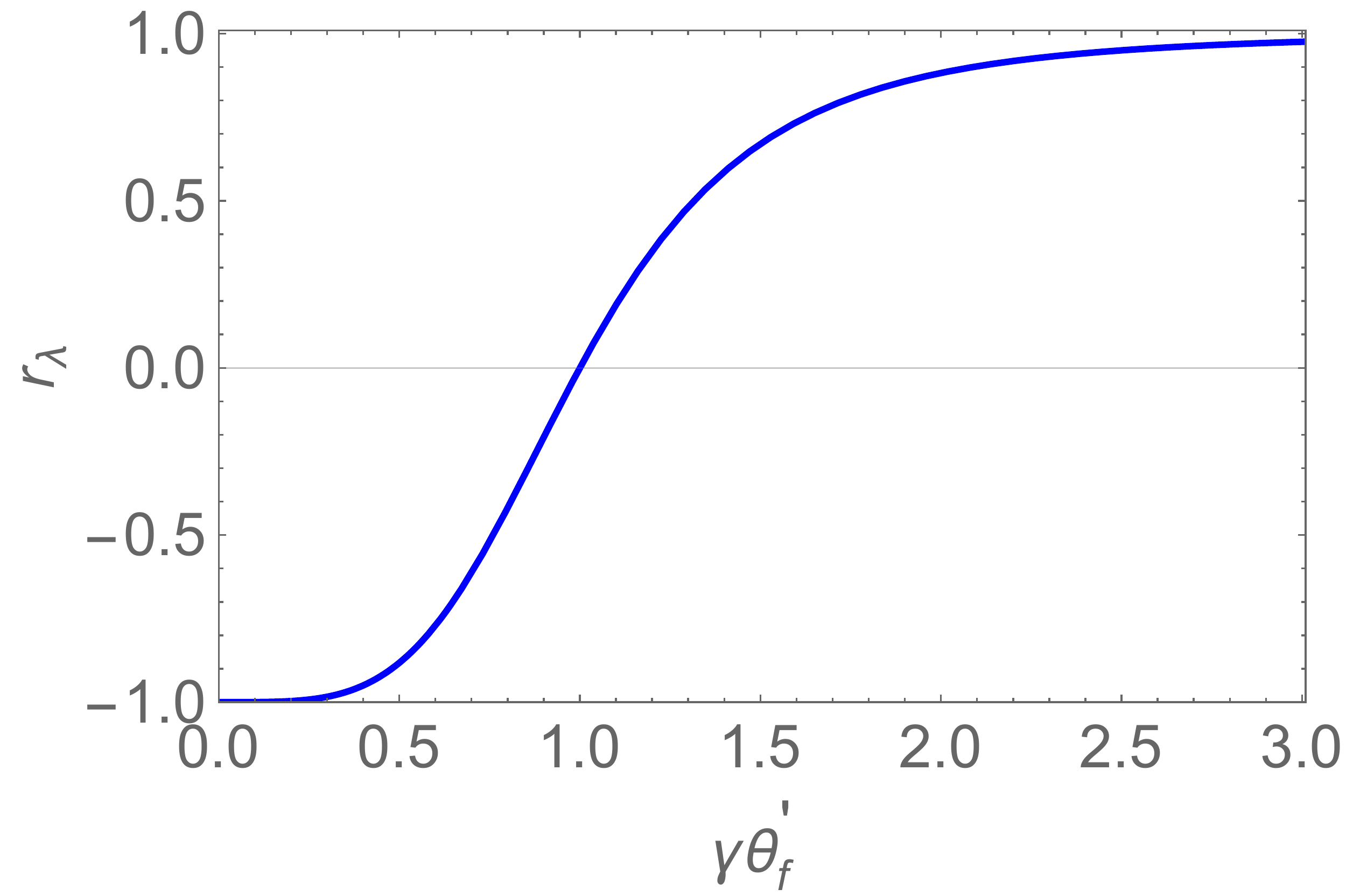}
	\caption{The ratio $r_\lambda = \langle \lambda_f \rangle/\lambda_i$ of the mean helicity (\ref{eq:mean_helicity}) of the outgoing photon to the incident--photon helicity $\lambda_i$, obtained for the $n S_{0} \to n' P_{1} \to n S_{0}$ resonant transition. The result is presented in the collider frame.}
    \label{Fig3}
\end{figure}
%
% --------------------------------------------------------------------------------------------------
%

So far we have focused on the angular distribution of scattered x--rays as observed both in the projectile and the laboratory frames. However, Eqs.~(\ref{eq:matrix_evaluated_E1_0_to_0}) and (\ref{eq:diff_cross_section_0_to_0}) can be also employed to investigate the \textit{polarization transfer} between incident and outgoing photons. As mentioned already above, this can be done most easily if the polarization states are described within the helicity basis ${\bm \epsilon}_{\lambda}$ with $\lambda = \pm 1$. For example, by employing Eq.~(\ref{eq:polarization_vectore_helicity_basis}) the (polarization--sensitive) cross section (\ref{eq:diff_cross_section_0_to_0}) can be written in this basis as:
\begin{eqnarray}
    \label{eq:diff_cross_section_0_to_0_circular}
    \frac{{\rm d}\sigma^{\rm (pl)}}{{\rm d}\Omega_f} &=&  \left|R_{J_\nu = 1}(\omega_i)\right|^2 \, \frac{1}{4} \, ( 1 - \lambda_i \lambda_f \cos\theta_{f} )^2 \, ,
\end{eqnarray}
where, similar to before, we assumed the head--on collisions. This expression immediately implies that the photon helicity flips, 
\begin{equation}
    \label{eq:helicity_transfer_0_0}
    \lambda_f = -\lambda_i, \;\;\mbox{when}\;\; \theta_i=\pi,\;\theta_f=0 \, ,
\end{equation}
i.\,e. for the strictly backward scattering. It can be well expected from the conservation of the total angular momentum projection of the entire system ``ion + light'' and by the fact that ionic state is not changed during the scattering process, $J_i = J_f =0$ and $M_i = M_f = 0$.

Even a small deviation from the strictly backward scattering leads to a failure of the relation (\ref{eq:helicity_transfer_0_0}). Indeed, the scattering at angles $0 < \theta_f < 180$ may proceed via both, helicity non--flip ($\lambda_f = + \lambda_i$) and helicity flip ($\lambda_f = - \lambda_i$) channels. The probability of these two channels can be easily obtained from the cross section (\ref{eq:diff_cross_section_0_to_0}) and reads as:
\begin{equation}
    \label{eq:helicity_flip_probability_0_0}
    W(\lambda_f = \pm \lambda_i) = \frac{(1 \mp \cos\theta_{f})^2}{2(1 +\cos^2\theta_{f})} \, .
\end{equation}
By using this expression one can obtain the \textit{mean} helicity of the final photon:
\begin{equation}
    \label{eq:mean_helicity}
     \langle \lambda_f \rangle = -\fr{2\cos\theta_f}{1+\cos^2\theta_f}\, \lambda_i\approx 
     -\frac{1-(\gamma \theta'_f)^4}{1+(\gamma \theta'_f)^4}\,\lambda_i \, ,
\end{equation}
that depends, of course, on the initial--state helicity $\lambda_i$ as well as on the scattering angle, see also Fig.~\ref{Fig3}. Note, that in Eq.~(\ref{eq:mean_helicity}) we indicate the angular dependence both in the ion--rest and the collider frame, where the latter case is considered for the experimentally relevant forward scattering regime, $\theta'_f \lesssim 1/\gamma$. \\

\subsubsection{$n S_{1/2} \to n' P_{1/2} \to n S_{1/2}$ scattering}

As the next example we consider the $n S_{1/2} \to n' P_{1/2} \to n S_{1/2}$ resonant scattering which is planned to be studied in the framework of the Gamma Factory proof--of--principle experiment \cite{BuC20}. For this case $J_i = J_f = J_\nu = 1/2$ and the summation over the index $k$ in the matrix element (\ref{eq:matrix_evaluated_E1}) runs from $k = 0$ to $k = 1$:
\begin{widetext}
\begin{eqnarray}
   \label{eq:matrix_evaluated_E1_2}
   {\cal M}^{({\rm pl}, E1)}_{M_f, M_i}(n S_{1/2} \to n' P_{1/2})
   && \nonumber\\[0.2cm]
   && \hspace*{-3cm} = - \frac{(4 \pi)^2 \alpha}{\sqrt{2}} \: S^{J_\nu = 1/2}_{E1, \,E1}(\omega_i) \, \sum\limits_{k q} (-1)^k \, \sqrt{2k + 1} \sprm{k q \, J_f M_f}{J_i M_i} \:
    \, \sixjm{1}{1}{k}{1/2}{1/2}{J_\nu} \, T_{k \, q}^{E1; E1}(\hat{\bm{k}}_i, \bm{\epsilon}_i \,; \hat{\bm{k}}_f, \bm{\epsilon}_f) \, \nonumber \\[0.2cm]
   && \hspace*{-3cm} = \pi \alpha S^{J_\nu = 1/2}_{E1, \,E1}(\omega_i) \, \left( (\bm{\epsilon}^*_{f} \cdot \bm{\epsilon}_{i}) \, - \, i \sqrt{3} \sum\limits_{q} \sprm{1 q \, 1/2 M_f}{1/2 M_i} \, \left[ \bm{\epsilon}^*_f \times \bm{\epsilon}_{i} \right]_q
   \right)    
   \, .
\end{eqnarray}
\end{widetext}
One can see that the matrix element ${\cal M}^{({\rm pl}, E1)}_{M_f, M_i}(n S_{1/2} \to n' P_{1/2})$ is written as a sum of two terms: the standard ``non--relativistic'' part $(\bm{\epsilon}^*_{f} \cdot \bm{\epsilon}_{i})$ and the second term, that is proportional to vector product of polarization vectors of incident and outgoing photons. This implies that---due to the fine structure of ionic levels---the application of simple (non--relativistic) formula (\ref{eq:matrix_evaluated_E1_0_to_0}) for the angular and polarization analysis of the $n S_{1/2} \to n' P_{1/2} \to n S_{1/2}$ transition is not valid. It can be also seen if we employ matrix element (\ref{eq:matrix_evaluated_E1_2}) to calculate the angle--differential scattering cross section:
\begin{eqnarray}
    \label{eq:diff_cross_section_12_to_12}
    &&\frac{{\rm d}\sigma^{\rm (pl)}}{{\rm d}\Omega_f} = 
    \frac{1}{2} \sum\limits_{M_i M_f}|{\cal M}^{({\rm pl}, E1)}_{M_f, M_i}(n S_{1/2} \to n' P_{1/2}) |^2 
    \nonumber \\[0.2cm]
    &&= \frac{1}{4}\left|R_{J_\nu = 1/2}(\omega_i)\right|^2
    \left[ 1 + \left| \left( {\bm \epsilon}^*_f \cdot {\bm \epsilon}_i \right) \right|^2 - \left| \left( {\bm \epsilon}_f \cdot {\bm \epsilon}_i \right) \right|^2 \right],
\end{eqnarray}
where the summation over the final-- and averaging over initial ionic substates is performed. Moreover, in the second line of this expression we used Eq.~(\ref{eq:R_function_definition}) and the relation $\sum_{q} |[\bm{\epsilon}^*_f \times \bm{\epsilon}_{i}]_q |^2 = 1 - \left|(\bm{\epsilon}_f \cdot \bm{\epsilon}_{i})\right|^2$, c.f.~Ref.~\cite{VaM88}. 

The differential cross section (\ref{eq:diff_cross_section_12_to_12}) for the $n S_{1/2} \to n' P_{1/2} \to n S_{1/2}$ obviously differs from Eq.~(\ref{eq:diff_cross_section_0_to_0}) that describes both, the relativistic $n S_{0} \to n' P_{1} \to n S_{0}$ and non--relativistic $ns \to n'p \to ns$ transitions. This difference becomes even more remarkable if we average over the initial-- and sum over the final polarization states of photons:
\begin{eqnarray}
    \label{eq:diff_cross_section_12_to_12_unp_prrojectile}
    \frac{{\rm d}\sigma^{\rm (pl, unp)}}{{\rm d}\Omega_f} &=& \frac{1}{2}\left|R_{J_\nu = 1/2}(\omega_i)\right|^2.
\end{eqnarray}
This formula illustrates the well--known fact that the (polarization--averaged) angular distribution of the photons, scattered in the $n S_{1/2} \to n' P_{1/2} \to n S_{1/2}$ transition, is \textit{isotropic} in the ion rest frame, in contrast to the $1 + \cos^2\theta_f$ emission patter of the $ns \to n'p \to ns$ photons, c.f.~Eq.~(\ref{eq:diff_cross_section_0_to_0_unpolarized}). 

Similar to before, we can use Eq.~(\ref{eq:diff_cross_section_12_to_12_unp_prrojectile}) to obtain the total cross section:
\begin{equation}
    \label{eq:total_cross_section_12_12}
    \sigma^{\rm (pl, unp)} = 2 \pi \left|R_{J_\nu = 1/2}(\omega_i)\right|^2 \, ,
\end{equation}
and the angle--differential cross section in the laboratory frame:
\begin{equation}
    \label{eq:diff_cross_section_12_to_12_unp_laboratory}
    \frac{{\rm d}\sigma^{\rm (pl, unp)}}{{\rm d}\Omega'_f} = \left|R_{J_\nu = 1/2}(\omega_i)\right|^2 \,  \frac{2\gamma^2}{\left[1+(\gamma \theta'_f)^2\right]^2} \, .
\end{equation}
Due to the Lorentz transformation, the latter is anisotropic and predicts that most of the scattered photons is emitted in the forward direction.

\subsubsection{$n S_{1/2} \to n' P_{3/2} \to n S_{1/2}$ scattering}

As the last scenario, we briefly mention here the $n S_{1/2} \to n' P_{3/2} \to n S_{1/2}$ resonance scattering, whose angle--differential cross section reads---in the ion rest frame---as: 
\begin{eqnarray}
   \label{eq:diff_cross_section_polarization_1_3}
   \frac{d\sigma^{\rm (pl)}}{d\Omega_f} &=& \frac{1}{4}\left|R_{J_\nu = 3/2}(\omega_i)\right|^2 \nonumber \\
   &\times& \left[ \left| \left( {\bm \epsilon}^*_f \cdot {\bm \epsilon}_i \right) \right|^2 +\frac{1}{4}
   \left(1 - \left| \left( {\bm \epsilon}_f \cdot {\bm \epsilon}_i \right) \right|^2 \right) \right] \, .
\end{eqnarray}
Again, in order to obtain this expression we have averaged over initial-- and summed over final ionic substates. Moreover, if we additionally average (sum) over the polarization of the incident and outgoing photons and assume head--on collisions, $\theta_i = \pi$, we obtain: 
\begin{eqnarray}
   \label{eq:diff_cross_section_polarization_1_3_unp}
   \frac{d\sigma^{\rm (pl, unp)}}{d\Omega_f} &=& \frac{1}{32}\left|R_{J_\nu = 3/2}(\omega_i)\right|^2 \,
   \left(7 + 3\cos^2\theta_f \right) \, ,
\end{eqnarray}
which, upon integration over the photon scattering angle, gives the total cross section,
\begin{equation}
   \sigma^{\rm (pl, unp)} =  \pi \left|R_{J_\nu = 3/2}(\omega_i)\right|^2 \, ,
\end{equation}
for the unpolarized case.

\subsection{Effect of non--zero crossing angle}
\label{subsec:non_zero_angle_effect}

In the discussion above we have seen that the analysis of the angular distribution and polarization of scattered photons can be simplified if we assume that incident (primary) light propagates strictly antiparallel to the ion beam direction, $\theta_i = \pi$. However, this head--on collision scenario will not be realized at the Gamma Factory, where experiments are planned to be performed as small but non--vanishing crossing angles $\delta_i = \pi - \theta_i$ and, hence, for $\theta_i \ne \pi$. In the proof--of--principle experiment at the SPS facility, for example, the beam of lead Pb$^{79+}$ ions, moving with the Lorentz factor $\gamma=96.3$, will intersect the laser beam at the angle $\delta'_i=2^\circ$ in the collider frame, which corresponds to $\delta_i=2.3\times 10^{-4}$ in the ion--rest frame \cite{BuC20}. Even smaller crossing angles are expected for the collision experiments at the LHC, see Eq.~(\ref{eq:alpha_transformation_3}). 

In order to estimate how much non--zero $\delta_i$ may affect the angular and polarization properties of scattered photons we will re--visit the $n S_{0} \to n' P_{1} \to n S_{0}$ resonant scattering, whose angle--differential cross section is given by Eq.~(\ref{eq:diff_cross_section_0_to_0}). Again, for our analysis it is practical to express the polarization vectors of incident and outgoing photons in helicity basis, ${\bm \epsilon}_i = {\bm \epsilon}_{\lambda_i}$ and ${\bm \epsilon}_f = {\bm \epsilon}_{\lambda_f}$, and to write this cross section as:
\begin{eqnarray}
    \label{eq:diff_cross_section_0_to_0_non-zero_alpha}
    \frac{{\rm d}\sigma^{\rm (pl)}}{{\rm d}\Omega_f} &=& \left|R_{J_\nu = 1}(\omega_i)\right|^2 \nonumber \\
    &\times& \left(A_0 + A_1 \cos\phi + A_2 \cos2\phi \right) \, ,
\end{eqnarray}
where we considered the general case $\theta_i \ne \pi$ and, hence, $\delta_i \ne 0$. In this expression, moreover, $\phi=\varphi_f - \varphi_i$ and the coefficients $A_0$, $A_1$ and $A_2$ depend solely on helicities, $\lambda_i$ and $\lambda_f$, and polar angles, $\theta_i$ and $\theta_f$, of the photons:
\begin{subequations}
\begin{eqnarray}
    \label{eq:A_coefficients_0}
    A_0 &=& \frac{1}{4} \left(\cos\theta_i+ \lambda_i \lambda_f \cos\theta_f \right)^2 + 3 A_2 \,  , \\[0.2cm]
    A_1 &=& \frac{1}{8} \sin 2\theta_i \sin 2\theta_f +  \frac{\lambda_i \lambda_f}{2} \sin \theta_i \sin \theta_f  \,  , \\[0.2cm]
    A_2 &=& \frac{1}{8} \sin^2 \theta_i \sin^2 \theta_f \, .
     \label{eq:A_coefficients_2}
\end{eqnarray}
\end{subequations}

As seen from these expressions, the only coefficient
\begin{equation}
    A_0(\theta_f = \pi) = \frac{1}{4} \left(1 - \lambda_i \lambda_f \cos\theta_f \right)^2 \, 
\end{equation}
does not vanish for the head--on ion--photon collisions, for which $\theta_i = \pi$, thus resulting in the differential cross section (\ref{eq:diff_cross_section_0_to_0_circular}) that is independent on the azimuthal angle $\phi$. The departure from this head--on direction, i.\,e. when $\theta_i \ne \pi$ and $\delta_i \ne 0$, gives rise to all three \textit{non--zero} coefficients $A_0$, $A_1$ and $A_2$ and, hence, to the deviation from Eq.~(\ref{eq:diff_cross_section_0_to_0_circular}). However, for $\delta_i \ll 1$ these coefficients scale as:
\begin{equation}
    \label{eq:A_scaling}
     A_0 - A_0(\theta_f = \pi) \sim \delta_i^2, \;\; A_1\sim \delta_i,\;\; A_2\sim \delta^2_i,
\end{equation}
thus indicating that the effect of the non--zero crossing angle on the angle--differential cross section is of the order of $\delta_i$ and does not exceed 0.03~\% for the planned parameters of the Gamma Factory. This justifies the use of the (simplified) expressions for the head--on collisions of the laser photons and ion beams for the calculations of the angular distributions and polarization of scattered photons.

%
% -------------------------------- Section: Non-relativistic theory -----------------------------------
\section{Plane--wave scattering: Non--relativistic theory}
\label{sec:non-relativistic_theory}

\subsection{$1s \to 2p \to 1s$ scattering}
\label{subsec:non-rel_theory_resonant scattering}

Having discussed in detail the relativistic theory of the resonance photon scattering, we will briefly recall below the non--relativistic expressions. We restrict ourselves to the $1s \to 2p \to 1s$ transition in a hydrogen--like ion with the nuclear charge $Z$. The angle--differential cross section of this process can be obtained within the dipole approximation from Eq.~(63.1) of Ref.~\cite{BeL82} and reads as:
\begin{equation}
    \label{eq:diff_cross_section_1s_2p_non_rel}
    \frac{{\rm d}\sigma^{\rm (pl)}}{{\rm d}\Omega_f} = \frac{|V|^2\,\omega_i^4 }{(E_{\nu i}-\omega_i)^2+\Gamma_\nu^2/4} \, ,
\end{equation}
where the photon energy $\omega_i$ is again assumed to be close to the $E_{\nu i} = E_{2p} - E_{1s}$, and the natural width of the excited $2p$ state is given by:
\begin{equation}
    \label{eq:width_2p_non_relativistic}
    \Gamma_\nu = \left(\frac{2}{3}\right)^8 \; Z^4 \alpha^5 \, .
\end{equation}
Moreover, we introduce here the notation
\begin{eqnarray}
    \label{eq:V_matrix_element_non_rel}
    V &=& \sum_{m=0,\pm 1} ({\bm d}_{1s,2pm} \cdot {\bm \epsilon}_f^{*})({\bm d}_{2pm,1s} \cdot {\bm \epsilon_i}) \nonumber \\[0.2cm]
    &=& \frac{2^{15}}{3^{10} Z^2 \alpha} \, (\bm \epsilon_f^{*} \cdot {\bm \epsilon_i}) \, ,
\end{eqnarray}
for the sum over (the product of) matrix elements ${\bm d}_{1s,2pm} = \mem{1s}{{\bm d}}{2p \, m}$ of the electric dipole momentum $\bm d=-e \bm r$. By inserting the second line of this formula into Eq.~(\ref{eq:diff_cross_section_1s_2p_non_rel}) and making some simple algebra one can again obtain the angle--differential cross section (\ref{eq:diff_cross_section_0_to_0})--(\ref{eq:R_function_definition}) for the $1 S_{0} \to 2 P_{1} \to 1 S_{0}$ scattering. 

Finally, one can use Eq.~(\ref{eq:diff_cross_section_1s_2p_non_rel}) to evaluate the total scattering cross section and to determine its maximum values for the case when $\omega_i = E_{\nu i}$, i.e. for zero detuning,
\begin{eqnarray}
    \label{total-resonant-cross-section}
    \sigma^{\rm (pl, unp)}(\omega_i = E_{\nu i}) &=& \frac{3}{2\pi} \lambda^2_{\nu i} \nonumber \\[0.2cm]
    && \hspace*{-2cm} = (7.2 \times 10^{-19} \div 1.4 \times  10^{-13})\; {\rm cm}^2.
\end{eqnarray}
Here $\lambda_{\nu i} = 2 \pi /E_{\nu i}$ is the wave--length of the light, emitted (absorbed) in the $ns$--$n'p$ transition and the values in the second line are obtained for the ions, displayed in Table~\ref{tab1}. 

It is worth noting that the total cross section (\ref{total-resonant-cross-section}) can be obtained also from the fully relativistic expressions (\ref{eq:R_function_definition}) and (\ref{eq:total_cross_section_0_to_0_unpolarized}), where we assume that the branching ratio of the $ns$--$n'p$ transition is unity and, hence, $\Gamma_{\nu} = \Gamma_{\nu i}$. It is justified, therefore, to use this formula for the estimation of the scattering cross sections, which is important for the preparation of future Gamma Factory experiments. A more detailed analysis of the total cross sections will be presented in our forthcoming publication.

\subsection{Compton scattering}
\label{subsec:Compton}

While the present work is focused on the resonant scattering (\ref{eq:scheme_process}), it is informative to compare it with the Compton scattering of laser photons by ultra--relativistic electrons. Our interest to the Compton scattering stems from the fact that it is routinely used as a ``competitive'' source of intensive high--energy photon beams at accelerator facilities \cite{WeA09,NeT04}. Below we will briefly discuss the characteristics of this process, if observed at the parameters as the resonant scattering above.   

Similar to the resonant scattering, we consider the Compton process in the ion rest frame, where the incident photon energy $\omega_i \approx E_{\nu i}$ is much smaller than the electron rest--mass energy $511$ keV. It is justified, therefore, to describe the Compton effect within the non--relativistic approximation, whose discrepancy from the rigorous relativistic predictions does not not exceed 26~\% for the energies presented in Table~\ref{tab1}. The non--relativistic expression for the angle--differential Compton cross section can be found, for example, in the problem 14.12 of Ref.~\cite{GaK13} and reads as:
\begin{equation}
    \label{eq:Compton_angle_diff_cross_section}
    \frac{{\rm d} \sigma^{\rm (pl)}_{\rm C}}{{\rm d}\Omega_f} = r_e^2 \,
    \left|(\bm \epsilon_f^{*} \cdot {\bm \epsilon_i})\right|^2 \, .
\end{equation}
Here, $r_e$ is the classical electron radius and  ${\bm \epsilon}_i$ and ${\bm \epsilon}_f$ are the polarization vectors of the incident and outgoing photons. One immediately recognizes that the (non--relativistic) Compton matrix element in Eq.~(\ref{eq:Compton_angle_diff_cross_section}) has the same structure as the one (\ref{eq:matrix_evaluated_E1_0_to_0}) for the $nS_0 \to n'P_1 \to nS_0$ scattering. Therefore, the angular distribution of the Compton--scattered photons, 
\begin{equation}
    \label{eq:diff_cross_section_Compton}
    \frac{{\rm d} \sigma^{\rm (pl, unp)}_{\rm C}}{{\rm d}\Omega_f} = \frac{r_e^2}{2} \,(1+ \cos^2 \theta_f) \, ,
\end{equation}
obtained upon averaging over the polarizations of the initial-- and summing up over the polarizations of the final--state photons photons, resembles Eq.~(\ref{eq:diff_cross_section_0_to_0_unpolarized}). 

By performing the integration over the photon emission angles, we obtain from Eq.~(\ref{eq:diff_cross_section_Compton}) the well--known Thomson cross section:
\begin{equation}
    \label{eq:Thompson_cross_section}
    \sigma_{\rm Th} = \frac{8\pi}{3}\, r_e^2  = 6.7\times 10^{-25}\; {\rm cm}^2 \, ,
\end{equation}
which is \textit{six to twelve orders of magnitude} smaller than the resonant--scattering cross section (\ref{total-resonant-cross-section}). One can argue, therefore, that the Compton scattering will play virtually no role in the future Gamma Factory experiments.

%
% ------------------------------- Section: Scattering of twisted photons ----------------------------
%
\section{Scattering of twisted photons}
\label{sec:scattering_twisted}

\subsection{Angular distribution of scattered photons} 

Until now our analysis of the resonant scattering process (\ref{eq:scheme_process}) has been restricted to the scenario of incident plane--wave photons. Apart of this ``conventional'' radiation, the use of twisted (or vortex) light modes is also discussed in the framework of the Gamma Factory project \cite{BuC20}. In order to study the features of the scattering process for this second ``twisted'' case, we will consider below the head--on collisions between ions and incident photons, prepared in the so--called \textit{Bessel} state \cite{JaH05,ScF14,She17,KnS18}. During the recent years, the Bessel photon states have been widely employed in theoretical studies of various fundamental atomic processes \cite{AfC16,AfC18,AfC18b,ScF14,JeS11,JeS11b,AfC14,QuS19}. A great advantage of these states is that they allow to account both for paraxial and non--paraxial regimes, depending on the choice of kinematic parameters of the light. 

The Bessel photons are characterized by both, helicity $\lambda_i$ and the well--defined projection $m_i$ of the total angular momentum (TAM) onto their propagation direction, which is anti--parallel to the quantization ($z$--) axis in our case. Moreover, they possess energy $\omega_i$ as well as longitudinal $k_{iz}$ and (absolute value of) transverse $\varkappa_i = \left| {\bm \varkappa}_i \right|$ components of the linear momentum. As usual in the theory of twisted light, the ratio of the transverse to the longitudinal momenta of Bessel photons is parameterized in terms of the so--called opening angle. Since in our setup the twisted light counterpropagates the ion beam and, hence, $k_{iz} < 0$, this angle is defined as $\theta^{\rm (tw)}_k = \arctan\left(\varkappa_i/|k_{iz}|\right)$. Moreover, by recalling the standard representation of the Bessel states as a coherent \textit{superposition} of plane--wave components \cite{ScF14,JeS11}, one can trivially find that $\theta^{\rm (tw)}_k$ defines also the angle at which these plane waves cross the quantization axis. In what follows, therefore, the relation $\delta_i = \theta^{\rm (tw)}_k$ will be used to relate the scattering properties of twisted and plane--wave photons.   

Any theoretical analysis of the resonant scattering of Bessel photons in the state $\ketm{\varkappa_i, m_i, k_{iz}, \lambda_i}$ can be traced back to the second--order matrix element similar to that in Sec.~\ref{subsec:relativistic_matrix_element}. By using the standard approach, discussed in detail in Refs.~\cite{AfC16,KnS18}, this matrix element can be written in terms of its plane--wave conuterpartners: 
\begin{widetext}
\begin{eqnarray}
    \label{eq:matrix_element_twisted}
    {\cal M}^{\rm (tw)}_{M_f, M_i}({\bm b}) &=& i^{m_{i}} \int_0^{2\pi} 
    \frac{{\rm d}\varphi_i}{2\pi} \, {\rm e}^{-i m_i \varphi_i +  i \varkappa_i b \cos(\varphi_i-\varphi_b)} \, {\cal M}^{{\rm (pl)}}_{M_f, M_i} \, .
\end{eqnarray}
\end{widetext}
Here, ${\cal M}^{{\rm (pl)}}_{M_f, M_i} = {\cal M}^{{\rm (pl)}}_{M_f, M_i}(\theta_i, \varphi_i)$ is obtained from Eq.~(\ref{eq:amplitude_general}) for the plane--wave photons, incident with the momentum ${\bm k}_i = \left(\varkappa_i \cos\varphi_i, \varkappa_i \sin\varphi_i, k_{iz} \right)$ and with the non--zero crossing angle $\delta_i = \pi - \theta_i = \arctan\left(\varkappa_i/|k_{iz}|\right)$. In Eq.~(\ref{eq:matrix_element_twisted}) we introduced, moreover, the impact parameter $\bm b = (b \cos\varphi_b,\, b\sin\varphi_b, \,0)$ in order to specify the position of a target ion with respect to the vortex axis of the Bessel beam. It is also assumed that the scattered radiation is observed by the conventional (``twistedness''--insensitive) detectors and, hence, described the outgoing photons by the plane--wave state with particular wave--vector ${\bm k}_f$ and helicity $\lambda_f$.  

By making use of the matrix element (\ref{eq:matrix_element_twisted}) one can evaluate the angle--differential cross section for the resonant scattering of Bessel photons:
\begin{eqnarray}
    \label{eq:diff_cross_section_twisted_b_fixed}
    \frac{{\rm d}\sigma^{(\rm tw)}(\theta_f, \varphi_f; \, \delta_i, {\bm b})}{d\Omega_f} && \nonumber\\[0.2cm] 
    && \hspace*{-2cm} = \frac{1}{2J_i + 1} \sum\limits_{M_i M_f} \left| {\cal M}^{\rm (tw)}_{M_f, M_i}({\bm b}) \right|^2 \, ,
\end{eqnarray}
where we again performed averaging over the initial and summation over the final ionic states. We note that this cross section describes the scenario of a single ion, located at the impact parameter ${\bm b}$ with respect to the vortex axis. This illustrates the \textit{qualitative} difference between the interactions with plane--wave and twisted laser beams. The analysis of the latter case requires the knowledge about the spatial composition of the ionic target. In the Gamma Factory experiments, for example, the twisted laser photons will interact with counterpropagating ion beams having \textit{macroscopic} cross--sectional areas. Assuming for simplicity that ions are uniformly distributed over the entire transverse plane of such beams, we can derive from Eq.~(\ref{eq:diff_cross_section_twisted_b_fixed}) the ${\bm b}$--averaged cross section:
\begin{eqnarray}
    \label{eq:diff_cross_section_twisted_averaged}
    \frac{{\rm d}\bar\sigma^{(\rm tw)}(\theta_f, \varphi_f; \, \delta_i)}{d\Omega_f} &=&
    \int \frac{{\rm d}{\bm b}}{\pi R^2}\frac{{\rm d}\sigma^{(\rm tw)}(\theta_f, \varphi_f; \, \delta_i, {\bm b})}{d\Omega_f} \nonumber \\
    && \hspace*{-2cm} = \frac{1}{|\cos\delta_i|} \int\limits_0^{2\pi} \frac{{\rm d} \varphi_i}{2\pi}
    \frac{{\rm d}\sigma^{\rm (pl)}(\theta_f, \varphi_f; \, \delta_i, \varphi_i)}{{\rm d}\Omega_f} \, ,
\end{eqnarray}
where $R$ is the radius of the laser Bessel beam and ${\rm d}\sigma^{\rm (pl)}(\theta_f,\varphi_f; \, \delta_i , \varphi_i)/{\rm d}\Omega_f \equiv {\rm d}\sigma^{\rm (pl)}/{\rm d}\Omega_f$ is the conventional cross section for the scattering of incident plane--wave photons with the momentum ${\bm k}_i = \left(\varkappa_i \cos\varphi_i, \varkappa_i \sin\varphi_i, k_{iz} \right)$. We note that the second line of Eq.~(\ref{eq:diff_cross_section_twisted_averaged}) is obtained by following the standard averaging procedure as discussed in details in Refs.~\cite{ScF14,JeS11}.  

As seen from Eq.~(\ref{eq:diff_cross_section_twisted_averaged}), further evaluation of the differential cross section for the scattering of Bessel photons by macroscopic ionic target requires to define its plane--wave counterpartner ${\rm d}\sigma^{\rm (pl)}/{\rm d}\Omega_f$. By considering, for example, the $nS_0 \to n'P_1 \to nS_0$ transition, whose (plane--wave) cross section is given by Eq.~(\ref{eq:diff_cross_section_0_to_0_non-zero_alpha}), we obtain:
\begin{eqnarray}
    \label{eq:diff_cross_section_twisted_averaged_0_to_1}
    \frac{{\rm d}\bar\sigma^{(\rm tw)}(\theta_f; \, \delta_i)}{d\Omega_f} &=& \frac{\left|R_{J_\nu = 1}(\omega_i)\right|^2}{|\cos\delta_i|} \nonumber \\[0.2cm]
    && \hspace*{-2cm}  \times \int\limits_0^{2\pi} \frac{{\rm d} \varphi_i}{2\pi}
    \left(A_0 + A_1 \cos\phi + A_2 \cos2\phi \right) \nonumber \\[0.2cm]
    &=& \frac{\left|R_{J_\nu = 1}(\omega_i)\right|^2}{|\cos\delta_i|} A_0 \, ,
\end{eqnarray}
where the coefficients $A_0$, $A_1$ and $A_2$ are given by Eqs.~(\ref{eq:A_coefficients_0})--(\ref{eq:A_coefficients_2}). As seen from this expression, the angular distribution of scattered (plane--wave) photons is independent both, on the azimuthal angle $\varphi_f$ and the total angular momentum projection $m_i$ of incident Bessel beam. It remains still sensitive, however, to the helicities of incoming and going photons as follows from Eq.~(\ref{eq:A_coefficients_0}). By performing averaging (summation) over these helicities one finally obtains:
\begin{eqnarray}
    \label{eq:diff_cross_section_twisted_averaged_0_to_1_final}
    \frac{{\rm d}\bar\sigma^{(\rm tw, unp)}(\theta_f; \, \delta_i)}{d\Omega_f} &=& 
    \frac{\left|R_{J_\nu = 1}(\omega_i)\right|^2}{4 |\cos\delta_i|} \left[ 2+ 2\cos^2\delta_i\cos^2\theta_f\right. \nonumber \\[0.2cm]
    &+& \left. \sin^2\delta_i \sin^2\theta_f \right] \, , 
\end{eqnarray}
which restores the well--known plane--wave result (\ref{eq:diff_cross_section_0_to_0_unpolarized}) for the case when $\varkappa_i \to 0$ and, hence, $\delta_i = 0$. 

So far we have discussed the evaluation of the angle--differential cross section ${\rm d}\sigma^{\rm (tw)}/{\rm d}\Omega_f$ in the ion rest frame. However, for the planning and analysis of the future Gamma Factory experiments it is necessary to study the properties of the resonant scattering in the laboratory frame. The transformation between the frames for the incident twisted light requires some attention. For example, in order to induce the resonant transition (\ref{eq:scheme_process}) in ions, moving with the Lorentz factor $\gamma$, the counterpropagating Bessel photons with the energy
\begin{equation}
    \label{eq:energy_lab_frame}
    \omega_i' \approx \frac{E_{\nu} - E_i}{\gamma (1 + |\cos\delta'_i|)} = \frac{E_{\nu i}}{\gamma (1  + |\cos\delta'_i|)}
\end{equation}
in the \textit{collider} frame have to be employed. For non--zero opening angle $\delta'_i = \arctan(\varkappa'_i/|k'_{iz}|)$, $0 < \delta'_i < \pi/2$, this energy is larger than the energy of the plane--wave radiation for the head--on scenario, c.f.~Eq.~(\ref{eq:energy_initial_transformation}). We note, moreover, that the angle $\delta'_i$ is defined here also in the CF frame and, according to Eq.~(\ref{eq:alpha_transformation_2}), is about $2\gamma$ times larger than $\delta_i$ from the IRF frame.   

With the help of the relativistic kinematic relations from Sec.~\ref{sec:kinematic} one can also calculate the angle--differential cross section in the laboratory frame. In particular, for the realistic experimental scenario of the high--$\gamma$ collisions between Bessel laser-- and macroscopic ion beams we obtain:
\begin{eqnarray}
    \label{eq:diff_cross_section_twisted_averaged_lab_frame}
    \frac{{\rm d}\bar\sigma^{(\rm tw)}(\theta'_f, \varphi'_f; \delta'_i)}{d\Omega'_f} &\approx&
    \frac{4\gamma^2}{[1+(\gamma \theta'_f)^2]^2} \nonumber \\[0.2cm]
    &\times& \frac{{\rm d}\bar\sigma^{(\rm tw)}(\theta_f, \varphi_f; \delta_i)}{d\Omega'_f} \, ,
\end{eqnarray}
where $\varphi'_f = \varphi_f$, $\delta'_i \approx 2\gamma \delta_i $ and the polar emission angles $\theta_f$ and $\theta'_f$ are related to each other by Eq.~(\ref{eq:angle_transformation_exact}). Moreover, the cross section (\ref{eq:diff_cross_section_twisted_averaged}), as derived in the ion--rest frame, enters in the second line of this expression. Similar to before, we need to specify a particular transition in order to further evaluate Eq.~(\ref{eq:diff_cross_section_twisted_averaged_lab_frame}). By choosing again the $nS_0 \to n'P_1 \to nS_0$ transition and employing Eq.~(\ref{eq:diff_cross_section_twisted_averaged_0_to_1}) we obtain:
\begin{eqnarray}
    \label{eq:diff_cross_section_twisted_averaged_0_to_1_lab_frame}
    \frac{{\rm d}\bar\sigma^{(\rm tw)}(\theta'_f ; \delta'_i)}{d\Omega'_f} &=&
    \frac{\left|R_{J_\nu = 1}(\omega_i)\right|^2}{|\cos(\delta'_i/ 2\gamma)|}
    \frac{4\gamma^2 A_0}{[1+(\gamma \theta'_f)^2]^2}  \, ,
\end{eqnarray}
where $A_0$ is given by Eq.~(\ref{eq:A_coefficients_0}).

In the theoretical analysis above we have assumed the incident photon beam being in the well--defined Bessel state  $\ketm{m_i,\,\varkappa_i,\, k_{iz},\,\lambda_i}$ and counter--propagates the quantization ($z$--) axis. The angle--differential scattering cross sections, derived for this case both in the IRF and CF, are insensitive to the total angular momentum $m_i$ and the azimuthal angle $\varphi_f$ of scattered light. This behaviour can be easily understood from the symmetry consideration and has been discussed in the literature \cite{ScF14,KnS18,PeV18}. It is also known that the $\varphi_f$-- and $m_i$--dependence of the cross sections can be restored if the incident radiation is prepared as a coherent superposition of two Bessel states with different projections $m_{i,a}$ and $m_{i,b}$ of the total angular momentum:
\begin{eqnarray}
    \ketm{\varkappa_i,\, k_{iz},\,\lambda_i} &=& c_a \ketm{m_{i,a},\,\varkappa_i,\, k_{iz},\,\lambda_i} \nonumber \\[0.2cm]
    &+& c_b \ketm{m_{i,b},\,\varkappa_i,\, k_{iz},\,\lambda_i} \, ,
\end{eqnarray}
where $c_{a,b} = {\rm e}^{i \beta_{a,b}} \left|c_{a,b}\right|$ and $\left|c_a\right|^2 + \left|c_b\right|^2 = 1$. Indeed, if the incident light in the state $\ketm{\varkappa_i,\, k_{iz},\,\lambda_i}$ induces the $nS_0 \to n'P_1 \to nS_0$ transition, the angular distribution of outgoing photons is given in the collider frame by:
\begin{eqnarray}
    \label{eq:diff_cross_section_twisted_averaged_0_to_1_lab_frame_superposition}
    \frac{{\rm d}\bar\sigma^{(\rm tw)}(\theta'_f, \varphi'_f ; \delta'_i)}{d\Omega'_f} && \nonumber \\[0.2cm]
    && \hspace*{-2.3cm} =
    \frac{\left|R_{J_\nu = 1}(\omega_i)\right|^2}{|\cos(\delta'_i/ 2\gamma)|}
    \frac{4\gamma^2}{[1+(\gamma \theta'_f)^2]^2}  \nonumber \\[0.2cm]
    && \hspace*{-2.3cm} \times \left\{ A_0 + {\tilde A}_{\Delta m} \,\cos\left[\Delta m (\varphi_f-\pi/2)+\Delta\beta \right] \right\} \, ,
\end{eqnarray}
where $\Delta m = m_{i,b} - m_{i,a}$, $\Delta \beta = \beta_b - \beta_a$, and  
\begin{equation}
 {\tilde A}_{\Delta m}= |c_1 c_2|\times \left\{ \begin{array}{cl}
                                           A_1 & \mbox{at}\;\Delta m=\pm 1 \\
                                           A_2 & \mbox{at}\;\Delta m=\pm 2 \\
                                           0 & \mbox{at}\;\Delta m\neq \pm 1, \pm 2.
                                         \end{array}\right.
\end{equation}
Similar to the cross section (\ref{eq:diff_cross_section_twisted_averaged_0_to_1_lab_frame}), the parameters $A_0$, $A_1$ and $A_2$ are obtained here from Eqs.~(\ref{eq:A_coefficients_0})--(\ref{eq:A_coefficients_2}) with the angles $\delta_i = \pi -\theta_i$ and $\theta_f$ defined in the ion rest frame. Since in this frame $\delta_i <1 / \gamma$, the azimuthal anisotropy parameters are small, ${\tilde A}_{\Delta m = \pm 1} \sim \delta_i$ and ${\tilde A}_{\Delta m = \pm 2} \sim \delta^2_i$ and, hence, the $\varphi_f$-- and $m_i$--dependence of the scattering cross section will be strongly suppressed for the Gamma Factory scenario.

\subsection{TAM projection of the scattered photons}
\label{subsec:TAM_transfer}

In the previous Section we have discussed the experimental scenario, where incident light is twisted while the scattered photons are \textit{projected} to a plane--wave state $\ketm{{\bm k}_f \, \lambda_f}$, ``seen'' by a conventional detector. One can address a question, however, whether the outgoing radiation will be also twisted. This question attracts a considerable attention nowadays since it allows to understand the feasibility of the resonant backscattering by relativistic ions for the production of twisted gamma rays. In recent theoretical work by Tanaka and Sasao \cite{TaS21}, for example, the authors have projected the outgoing photons to the spherical waves to explore the ``twistendess'' transfer in the scattering process. In our present study, we propose an alternative approach and construct the matrix element for the process (\ref{eq:scheme_process}) in which incident photon in the Bessel state $\ketm{m_i, \, \varkappa_i, \, k_{iz},\, \lambda_i}$ counter--propagates the quantization axis, while the final--state photon $\ketm{m_f, \, \varkappa_f, \, k_{fz}, \, \lambda_f}$ is emitted along it. The matrix element of this ``twisted--to--twisted'' scattering can be again expressed: 
\begin{widetext}
\begin{eqnarray}
    \label{eq:matrix_element_twisted_twisted}
    {\cal M}^{\rm (tw-tw)}_{M_f, M_i}({\bm b}) &=& i^{m_f + m_i} \, \int_0 ^{2 \pi} \frac{{\rm d} \varphi_i}{2 \pi} \, \frac{{\rm d}\varphi_f}{2\pi} \, {\rm e}^{-i m_i \varphi_i +  i {\bm \varkappa}_i {\bm b} - i m_f \varphi_f - i {\bm \varkappa}_f {\bm b}} \, {\cal M}^{\rm (pl)}_{M_f, M_i} \, ,
\end{eqnarray}
\end{widetext}
in terms of the matrix element ${\cal M}^{\rm (pl)}_{M_f, M_i} = {\cal M}^{\rm (pl)}_{M_f, M_i}(\theta_i, \varphi_i)$ obtained for the plane--wave photons with the momenta ${\bm k}_i = \left(\varkappa_i \cos\varphi_i, \varkappa_i \sin\varphi_i, k_{iz} \right)$ and ${\bm k}_f = \left(\varkappa_f \cos\varphi_f, \varkappa_f \sin\varphi_f, k_{fz} \right)$. In this expression, moreover, ${\bm \varkappa}_{i,f} {\bm b} = \varkappa_{i,f} b \cos(\varphi_{i,f}- \varphi_b)$ is the scalar product of the transverse photon momentum ${\bm\varkappa}_{i,f}$ and the impact parameter ${\bm b}$ of a target ion. One can note that the same parameter ${\bm b}$ is used in ${\cal M}^{\rm (tw-tw)}_{M_f, M_i}({\bm b})$ to describe (the positions of) the axes of incident and outgoing photons. This simple choice is justified for the analysis of the collisions between ion and photon \textit{beams}, where the averaging over the $b$ has to be performed. A more general and, hence, more sophisticated scenario of shifted (with respect to each other) photon axes is out of scope of present paper and is discussed in Ref.~\cite{KaS21}.

In order to further evaluate the matrix element (\ref{eq:matrix_element_twisted_twisted}) we have to specify a particular transition and, hence, explicit form of the ${\cal M}^{\rm (pl)}_{M_f, M_i}$. Here we again consider the case of the $nS_0 \to n'P_1 \to nS_0$ resonant scattering whose plane--wave matrix element is given by Eqs.~(\ref{eq:T00_helicity})--(\ref{eq:matrix_evaluated_E1_0_to_0}). By using those expressions and by performing integration over the azimuthal angles $\varphi_{i}$ and $\varphi_{f}$ we obtain: 
\begin{eqnarray}
    \label{eq:matrix_element_twisted_twisted_0_1_0}
    {\cal M}^{({\rm tw-tw}, E1)}_{M_f = 0, M_i = 0}(n S_{0} \to n' P_{1}; {\bm b}) &=&  R_{J_\nu = 1}(\omega_i) \,  \nonumber \\[0.2cm]
    && \hspace*{-3.3cm} \times
    {\rm e}^{i(m_i + m_f) \varphi_b}
    \sum\limits_{\sigma} \Big[ J_{m_i+ \sigma} (\varkappa_i b) \, J_{m_f-\sigma} (\varkappa_f b) \nonumber \\
    && \hspace*{-3.3cm} \times d^{\;1}_{\sigma\lambda_i}(\theta_i) \, d^{\;1}_{\sigma\lambda_f}(\theta_f) \Big] \, .
\end{eqnarray}
This expression implies the complete transfer of the TAM projection between incident and outgoing photons:
\begin{equation}
    \label{eq:TAM_transfer}
    m_f =- m_i = 0, \pm 1 \, ,
\end{equation}
for the case of a target ion, located at the vortex line of the laser twisted beam, $b = 0$. This simple result is a consequence of the conservation of the TAM projection in the system ``ion + light'', that possesses axial symmetry for $b = 0$ and incident (outgoing) photons propagating anti--parallel (parallel) to the quantization axis. Moreover, as seen from Eq.~(\ref{eq:TAM_transfer}), the electric--dipole selection rules, valid for the $nS_0 \to n'P_1 \to nS_0$ transition, restrict the TAM projection of absorbed (or emitted) photon to $0, \, \pm 1$ only.  

From the fact that the plane--wave matrix elements of the  resonant-- $nS_0 \to n'P_1 \to nS_0$ and Compton scattering have the same structure, c.~f.~Sec.~\ref{subsec:Compton}, one can expect that the latter process also may result in the complete TAM transfer between incoming and outgoing photons. Indeed, the relation $m_f =- m_i$ has been predicted recently for the Compton \textit{backscattering} of twisted light off relativistic free electrons \cite{JeS11, IvS11}. 

\begin{figure}[t]
	\includegraphics[width=0.99\linewidth]{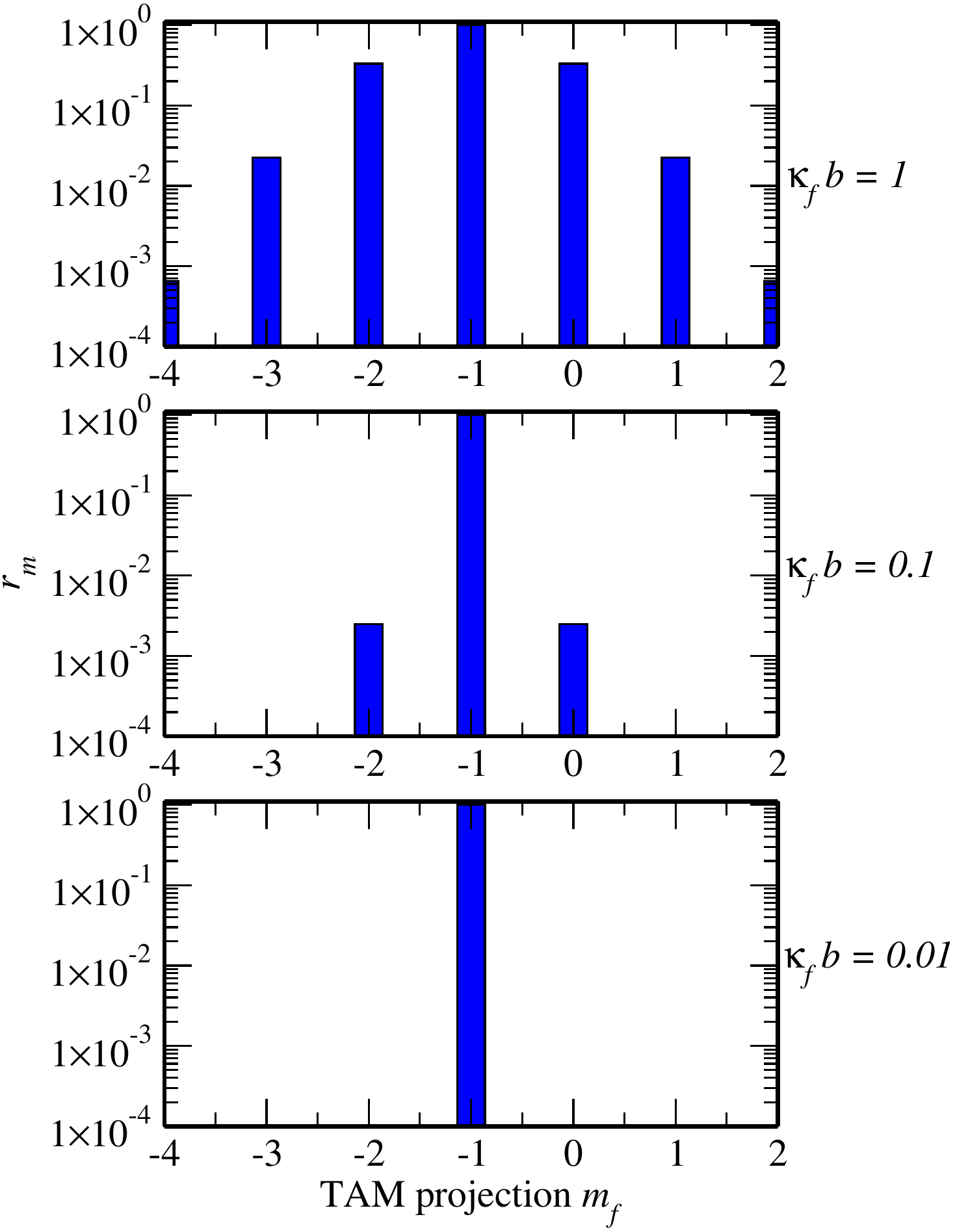}
	\caption{The ratio $r_m = \left|{\cal M}(m_f) \right|^2/ \left|{\cal M}(m_f = -m_i)\right|^2$ of the (squares of) matrix elements (\ref{eq:matrix_element_twisted_twisted_0_1_0_near_center}), that describe the backscattering of a photon with the TAM projection $m_f$. The calculations are performed for the incident photon with $m_i = 1$ and for the various values of the parameter $\varkappa_f b$. 
	}
    \label{Fig4}
\end{figure}

While the TAM--transfer relation (\ref{eq:TAM_transfer}) is derived for $b = 0$, one can see from Eq.~(\ref{eq:matrix_element_twisted_twisted_0_1_0}) that it will remain valid---at least approximately---also for small displacements of an ion from the beam axis, $0 < b \ll 1/\varkappa_i$. For this case the scattering matrix element can be simplified to: 
\begin{eqnarray}
    \label{eq:matrix_element_twisted_twisted_0_1_0_near_center}
    {\cal M}^{({\rm tw-tw}, E1)}_{0, 0}(n S_{0} \to n' P_{1}; {\bm b}) &\approx&  R_{J_\nu = 1}(\omega_i) \,  {\rm e}^{i(m_i + m_f) \varphi_b} \nonumber \\[0.2cm]
    && \hspace*{-3.5cm} \times
    J_{m_f + m_i} (\varkappa_f b) \, d^{\;1}_{-m_i \lambda_i}(\theta_i) \, d^{\;1}_{-m_i \lambda_f}(\theta_f) \, ,
\end{eqnarray}
which implies that the distribution of the TAM projections of the outgoing electron is peaked around $m_f = -m_i$ and the width of the peak becomes narrower as the factor $b \varkappa_f$ decreases. Indeed, it can be seen from the well--known asymptotic expression of the Bessel function: 
\begin{equation}
    J_{n} (x) \approx \sqrt{\frac{n}{3\pi x\,z}}\,e^{-z},
    \; {\rm when}\; z=\sqrt{\frac{8n^3}{9x}} \gg 1 \, ,
\end{equation}
and is illustrated in Fig.~.\ref{Fig4}. In contrast, for large impact parameters $b > 1/\varkappa_i$, the breakdown of the axial symmetry of the system ``ion + light'' will become significant and Eq.~(\ref{eq:TAM_transfer}) will not be valid anymore. For this latter case, the resonant photon scattering will result in emission of photons with TAM projections \textit{distributed} within a wide range of $ m_f$.

The transfer of the TAM projection between incoming and outgoing photons has been discussed so far only for the $nS_0 \to n'P_1 \to nS_0$ resonant transition. However, Eq.~(\ref{eq:matrix_element_twisted_twisted}) and the corresponding plane--wave matrix elements can be employed also to analyze the other two transitions, $n S_{1/2} \to n' P_{1/2, 3/2} \to n S_{1/2}$, discussed in the present work. Since the theoretical analysis for these two cases is rather similar to that of the $nS_0 \to n'P_1 \to nS_0$ scattering, we will not display it here and just briefly mention its most important results. In particular, for an ion located at the laser vortex line, $b = 0$, we find:
\begin{equation}
    \label{eq:TAM_transfer_2}
    m_f + M_f = -m_i + M_i \,  ,
\end{equation}
which implies that complete TAM transfer between incident and scattered photons is not anymore possible due to the spin--flip ionic transitions, $M_i \ne M_f$. Moreover, significant violation of this TAM selection rule has been again found for for large impact parameters $b > 1/\varkappa_i$. 

%
% ------------------------------------ Figure 4 -------------------------------------------
%
%
% ------------------------------------------------------------------------------------------
%

%
%
% ------------------------------ Section: Summary and outlook ------------------------------------
%
\section{Summary and outlook}
\label{sec:summary}

In summary, we have performed a theoretical investigation of the resonant elastic photon scattering by partially stripped heavy ions, moving with ultra--relativistic velocities. Special emphasis in our work was placed on the angular distribution and polarization of scattered photons as observed both in the collider and ion--rest frames. We have shown that these (angular and polarization) properties are characterized most naturally in terms of the so--called \textit{polarization tensor} that was derived upon analysis of the relativistic second--order matrix element for the scattering process. The great advantage of the polarization tensor formalism is that it provides a simple and elegant tool for the analysis of the polarization and even orbital--angular--momentum transfer between incident and outgoing radiation. 

While the derived formalism is general and can be used to investigate the resonant photon scattering by arbitrary (many--electron) ion, independent on its shell structure, we have focused here on the $n S_{0} \to n' P_{1} \to n S_{0}$ as well as $n S_{1/2} \to n' P_{1/2, 3/2} \to n S_{1/2}$ transitions, induced by conventional plane--wave radiation. A particular interest to these electric dipole ($E1$) transitions arise from the fact that are likely to be studied at planned Gamma Factory experiments. In order to support the future experiments, the angle--differential cross sections were derived for all three transitions and for two scenarious in which the polarization states of initial-- and final--state photons (i) are defined, or (ii) remain unobserved. The former scenario allowed us to investigate how the angular distribution and polarization of scattered radiation is affected if incident light is itself polarized.     

In the present work the polarization--tensor formalism has been used to analyze the resonant scattering of not only plane--wave but also twisted (or vortex) photons, which also attract particular attention in the Gamma Factory project. Owing to their non--trivial internal structure, twisted photon beams may introduce new features in the scattering process. We argued, in particular, that the photon energy in the ion rest frame can be fine--tuned by varying the opening angle of the vortex beam in the CF. This opening angle, which is usually employed to characterize the ratio of transverse to longitudinal momenta of twisted photons, can also affect the angular distribution of scattered photons. Furthermore, if incident photons are prepared in the superposition of two twisted states with different projections of their total angular momenta, the sensitivity of the angle--differential cross section to these projections can be also observed in the future Gamma Factory experiments, performed at moderate $\gamma$'s.

Yet another challenging question, that has been addressed with the help of the polarization tensor formalism, was whether twistedness is transferred between incident and outgoing photons in course of the scattering process. We have found that such a ``twistedness transfer'' critically depends on the spatial position of each particular ion. If, for example, an ion is located near the vortex axis of twisted light beam and undergoes $n S_{0} \to n' P_{1} \to n S_{0}$ transition, the TAM projection of incident photon will be completely transferred for the back--scattered final one. However, it is not a case for ions, displaced from the center of light beam, for which the scattered radiation will exhibit some distribution over the TAM projections.   

The theoretical analysis of the TAM, or ``twistedness'', transfer was performed in the present work for the simplest scenario of the head--on collisions between incident Bessel--photon and ion beams and for the case when the scattered light propagates along the $z$--axis of the system. The investigation of a more complex geometrical setup, in which the directions of incoming and outgoing twisted photons are \textit{not} exactly (anti--)parallel to the ion beam axis, is a rather demanding task, especially if carried out in the collider frame. This analysis would require, moreover, to treat twisted photons as spatially localized wave packets, similar to what has been done in Ref.~\cite{IvS11} for the description of the Compton scattering process. Such a theoretical study is out of scope of the present work and will be discussed in detail in a forthcoming publication.

\section*{Acknowledgements}

We are very grateful to D.~Karlovets, N.~Muchnoi, M.~W.~Krasny and D.~Budker for many fruitful discussions. This work was funded by the Deutsche Forschungsgemeinschaft (DFG, German Research Foundation) under Germany's Excellence Strategy EXC-2123, QuantumFrontiers 390837967. A.V. acknowledges financial support by the Government of the Russian Federation through the ITMO Fellowship and Professorship Program. Some work in Sec.~\ref{sec:scattering_twisted}  is supported by the Russian Science Foundation (Project No. 21-42-04412) and by the Deutsche Forschungsgemeinschaft (Project No. SU 658/5-1)

%\bibliography{./b.20.photon_scattering_GF}
%
%%%%%%%%%н%%%%%%%%%%%%%%%%%%%%н%%%%%%%%%н%%%%%%%%%%%%%%%%%%%%%%%%%%%%%%н%%%%
%%%%%%%%%н%%%%%%%%%%%%%%%%%%%%н%%%%%%%%%н%%%%%%%%%%%%%%%%%%%%%%%%%%%%%%н%%%%
\end{document}